%% file: ms.tex
\documentclass[useAMS,usenatbib,usegraphicx]{mn2e}

\newcommand{\Swift}{\textit{Swift}}

\newcommand{\fermi}{\textit{Fermi}}

\newcommand{\hostl}{SDSS\,J143840.98+373933.4}
\newcommand{\hosts}{SDSS\,J1438}
\newcommand {\aap}{A\&A}
\newcommand {\apj}{ApJ}
\newcommand {\apjl}{ApJL}
\newcommand {\apjs}{ApJS}
\newcommand {\aj}{AJ}
\newcommand {\nat}{Nat}
\newcommand {\mnras}{MNRAS}
\newcommand {\pasp}{PASP}

\newcommand {\araa}{ARA\&A}
\newcommand{\aapr}{A\&A~Rev.}
\newcommand{\rp}{\textit{r$^{\prime}$}}
\def\lesssim{\mathrel{\hbox{\rlap{\hbox{\lower3pt\hbox{$\sim$}}}\hbox{\raise2pt\hbox{$<$}}}}}
\def\gtrsim{\mathrel{\hbox{\rlap{\hbox{\lower3pt\hbox{$\sim$}}}\hbox{\raise2pt\hbox{$>$}}}}}
\newcommand\ion[2]{#1$\;${\small\rmfamily\@Roman{#2}}\relax}

\title{PTF10iya: A short-lived, luminous flare from the nuclear region of
a star-forming galaxy}

\author[Cenko et al.]{S. Bradley Cenko$^{1}$\thanks{E-mail: 
             \texttt{cenko@astro.berkeley.edu.}},
             Joshua S. Bloom$^{1}$,
             S. R. Kulkarni$^{2}$,
             Linda E. Strubbe$^{1,3}$,
\newauthor
             Adam A. Miller$^{1}$,
             Nathaniel R. Butler$^{1,4}$,
             Robert M. Quimby$^{5}$,
             Avishay Gal-Yam$^{6}$,
\newauthor
            Eran O. Ofek$^{2,4}$,
            Eliot Quataert$^{1,3}$,
            Lars Bildsten$^{7,8}$,
            Dovi Poznanski$^{4,9,1,10}$,
\newauthor
            Daniel A. Perley$^{1}$,
            Adam N. Morgan$^{1}$,
            Alexei V. Filippenko$^{1}$,
            Dale A. Frail$^{11}$,
\newauthor
            Iair Arcavi$^{6}$,
            Sagi Ben-Ami$^{6}$,
            Antonio Cucchiara$^{9,1}$,
            Christopher D. Fassnacht$^{12}$,
\newauthor
            Yoav Green$^{6}$,
            Isobel M. Hook$^{13,14}$,
            D. Andrew Howell$^{15,7}$,
            David J. Lagattuta$^{12}$,
\newauthor
            Nicholas M. Law$^{16}$,
            Mansi M. Kasliwal$^{2}$,
            Peter E. Nugent$^{9}$,
            Jeffrey M. Silverman$^{1}$,
\newauthor
            Mark Sullivan$^{13}$,
            Shriharsh P. Tendulkar$^{2}$,
            and Ofer Yaron$^{6}$. \\
$^{1}$Department of Astronomy,
  University of California, Berkeley, CA 94720-3411, USA \\
$^{2}$Cahill Center for Astrophysics, California Institute
  of Technology, Pasadena, CA, 91125, USA \\
$^{3}$Theoretical Astrophysics Center, University of
  California, Berkeley, CA 94720-3411, USA \\
$^{4}$Einstein Fellow \\
$^{5}$IPMU, University of Tokyo, Kashiwanoha 5-1-5, Kashiwa-shi, Chiba, 
  Japan \\
$^{6}$Department of Particle Physics and Astrophysics,
  The Weizmann Institute of Science, Rehovot 76100, Israel \\
$^{7}$Department of Physics, Broida Hall, University of
  California, Santa Barbara, CA 93106, USA \\
$^{8}$Kavli Institute for Theoretical Physics, Kohn Hall,
  University of California, Santa Barbara, CA 93106, USA \\
$^{9}$Computational Cosmology Center, Lawrence Berkeley
  National Laboratory, 1 Cyclotron Road, Berkeley, CA 94720, USA \\
$^{10}$School of Physics and Astronomy, Tel-Aviv University, 
  Tel-Aviv 69978, Israel \\
$^{11}$National Radio Astronomy Observatory, P.O. Box 0, Socorro, NM
  87801, USA \\
$^{12}$Department of Physics, University of California
  Davis, 1 Shields Avenue, Davis, CA 95616, USA \\
$^{13}$Department of Physics (Astrophysics), University of Oxford, 
  Keble Road, Oxford, OX1 3RH UK \\
$^{14}$INAF-Osservatorio di Roma, via Frascati 33, I-00040
  Monteporzio Catone (Roma), Italy \\
$^{15}$Las Cumbres Observatory Global Telescope Network,
  Goleta, CA 93117, USA \\
$^{16}$Dunlap Institute for Astronomy and Astrophysics,
  University of Toronto, 50 St. George Street, Toronto M5S 3H4,
  Ontario, Canada
}

\begin{document}





\pagerange{\pageref{firstpage}--\pageref{lastpage}} \pubyear{2011}

\maketitle

\label{firstpage}

\begin{abstract}
We present the discovery and characterisation of PTF10iya, a
short-lived ($\Delta t \approx 10$\,d, with an optical decay rate of 
$\sim 0.3$\,mag\,d$^{-1}$), luminous ($M_{g^{\prime}}
\approx -21$\,mag) transient source found by the Palomar Transient Factory.
The ultraviolet/optical spectral energy distribution
is reasonably well fit by a blackbody with $T \approx$ (1--2)  $\times 
10^{4}$\,K and peak bolometric luminosity $L_{\mathrm{BB}} \approx$
(1--5) $\times 10^{44}$\,erg\,s$^{-1}$ (depending on the details of the
extinction correction).  A comparable amount of energy is radiated 
in the X-ray band that appears to result from a distinct physical process.  
The location of PTF10iya is consistent with the nucleus of a
star-forming galaxy ($z = 0.22405 \pm 0.00006$) to 
within 350\,mas (99.7 per cent confidence radius), or a projected 
distance of less than 1.2\,kpc.  At first glance, these properties appear
reminiscent of the characteristic ``big blue bump'' seen in the 
near-ultraviolet spectra of many active galactic nuclei (AGNs).  However, 
emission-line diagnostics of the host galaxy, along with a 
historical light curve extending back to 2007, show no evidence 
for AGN-like activity.  We therefore consider whether the tidal disruption
of a star by an otherwise quiescent supermassive black hole may account 
for our observations.  Though with limited
temporal information, PTF10iya appears broadly consistent with the
predictions for the early ``super-Eddington'' phase of a 
solar-type star being disrupted by a $\sim 10^{7}$\,M$_{\odot}$ black hole.  
Regardless of the precise physical origin of the accreting material, 
the large luminosity and short duration suggest that otherwise 
quiescent galaxies can transition extremely rapidly to radiate 
near the Eddington limit; many such outbursts may have 
been missed by previous surveys lacking sufficient cadence. 
\end{abstract}


\section{Introduction}
\label{sec:intro}
Understanding the physics of accretion, and, in
particular, the associated electromagnetic emission, is a topic at the
forefront of modern astrophysics.  From planet formation to galactic
X-ray binaries to the most luminous quasars, accretion plays a role in
an astounding array of phenomena across a diverse range of mass and
size scales.  Of special interest is the importance of the process
in the formation and growth of the supermassive black holes (SMBHs)
that appear to reside in the centres of all bulge galaxies
\citep{kr95}, and the mechanism by which these SMBHs are intimately
connected to the growth and evolution of galaxies (i.e., the
$M_{\mathrm{BH}}$--$\sigma_{*}$ relation; \citealt{fm00,gbb+00}).

\citet{r88} first suggested that the tidal disruption of a star by a
SMBH could be a powerful method to study accretion
in otherwise quiescent galaxies.  If the observational
signature of the resulting electromagnetic emission, known as a tidal
disruption flare (TDF), were better constrained, these outbursts could
potentially be used to probe the properties (i.e., mass) of SMBHs in
quiescent galaxies beyond the reach of (resolved) kinematic studies of
central gas and stars.  

The task of observationally identifying TDFs, however, is greatly
complicated by the diverse array of transient phenomena that occur in
galactic nuclei.  Aside from the well-known classes of active galactic
nuclei (AGNs), such as blazars and normal Seyfert galaxies, potential 
TDF ``impostors'' include the relatively rare double-peaked emitters 
(systems that exhibit both blueshifted and redshifted emission lines that
may indicate an origin in a rotating accretion disc; \citealt{hf88}),
and Type IIn supernovae (SNe~IIn; objects with narrow and intermediate
width emission lines indicative of interaction with a moderately dense
circumstellar medium; see \citealt{f97} for a review).  Future
progress requires a full accounting of these interlopers, both at the
individual (i.e., to establish or rule out a TDF origin for a given
event) and class (i.e., for rate calculations) levels
\citep{sq10,vfg+10}.  

In the optical bandpass, wide-field, high-cadence
surveys such as the Palomar Transient Factory (PTF;
\citealt{lkd+09,rkl+09}), Pan-STARRS1 (PS1; \citealt{kab+02b}), the 
Catalina Real-time Transient Survey (CRTS; \citealt{ddm+09}), and
SkyMapper \citep{skf+05} are all currently attempting to chart the
bright end of the transient landscape in advance of the Large Synoptic
Survey Telescope (LSST).  As untargeted transient surveys, these
projects should provide a relatively unbiased view (with the notable
exception of dust extinction) of the optical variability of nearby
galaxies, where many AGNs and SNe emit a significant fraction of their
bolometric luminosity, and are therefore well suited to address many
of the questions mentioned above.

Here we present the discovery by PTF of a short-lived, luminous transient 
(PTF10iya) in the nuclear region of a redshift $z = 0.22$ star-forming
galaxy, which serves as an interesting test case for many of these
issues.  Our manuscript is organised as follows. In \S\ref{sec:obs},
we describe the discovery of PTF10iya, as well as 
optical, ultraviolet (UV), near-infrared (NIR), and X-ray follow-up
observations and archival data at the location of the transient.
Section \ref{sec:analysis} presents the astrometry of the transient
emission, the broadband spectral energy distribution, and the
properties of the host galaxy. In \S\ref{sec:interp}, we outline possible
emission mechanisms to explain the observed properties of the
outburst.  

Throughout this paper, we adopt a standard $\Lambda$CDM cosmology with
$H_{0}$ = 71\,km s$^{-1}$ Mpc$^{-1}$, $\Omega_{\mathrm{m}} = 0.27$, and
$\Omega_{\Lambda} = 1 - \Omega_{\mathrm{m}} = 0.73$ \citep{sbd+07}.  
 All quoted uncertainties are 1$\sigma$ (68\%) confidence
intervals unless otherwise noted, and UT times are used throughout.

\begin{figure*}
  \includegraphics[width=18cm]{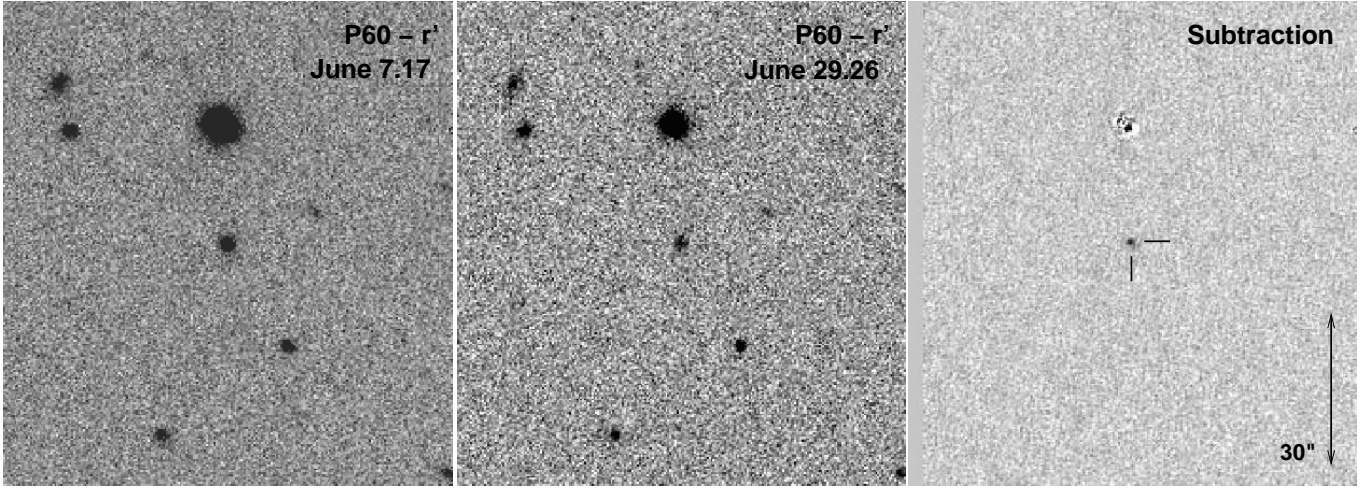}
  \caption{Finder chart for PTF10iya.  Left: P60 \rp\ image from
    2010 June 7.  Middle: P60 reference image from
    several weeks later.  Right: Digital subtraction of the two
    P60 images.  The transient PTF10iya is indicated by the black tick
    marks.  All images are oriented with north up and east to the left.}
\label{fig:finder1}
\end{figure*}

\section{Observations}
\label{sec:obs}

\subsection{Palomar 48-inch discovery and photometry}
\label{sec:obs:p48}
As part of the PTF 5-day cadence survey, we obtained a pair of $R$-band
 images of PTF field 4328 on 2010 June 6 with the Palomar 48-inch 
telescope (P48) equipped with the refurbished CFHT12k camera
\citep{rsv+08}.  Subtraction of a stacked reference image 
of the field with HOTPANTS\footnote{See
  http://www.astro.washington.edu/users/becker/hotpants.html.} 
revealed a new transient source at coordinates $\alpha
= 14^{\mathrm{h}} 38^{\mathrm{m}} 41.00^{\mathrm{s}}$, $\delta = 37^{\circ}
39\arcmin 33\farcs6$ (J2000.0), with an astrometric 
uncertainty of $\pm 150$\,mas in each coordinate 
(referenced with respect to the Sloan Digital Sky Survey Data Release
7 catalog; \citealt{aaa+09e}).  

The transient was discovered three hours later by Oarical, an 
autonomous software framework of the PTF collaboration 
\citep{brn+11}, and given the name PTF10iya.  The software further
noted the presence of a coincident catalogued extended source, \hostl\
(hereafter \hosts; Figure~\ref{fig:finder1}), identifying this object
as a potential host galaxy for PTF10iya.  

No transient emission was detected at this location with P48, either prior
to (extending back to 2009 May) or after (through 2011 July)
the outburst, to a typical 3$\sigma$ limiting magnitude of $R \approx 21$.
A listing of P48 observations taken around the time of
outburst, calibrated using Sloan Digital Sky Survey 
(SDSS) magnitudes of nearby point sources and
the filter transformations of \citet{jga06}, is provided in
Table~\ref{tab:optuv}.  

This field was also observed as part of the Palomar-QUEST survey
\citep{dbm+08} on 8 separate nights ranging from 2007 May 6 to
2008 July 30.  The observations have been compiled into a single
searchable database at Lawrence Berkeley National Laboratory as part
of the Deep Sky project\footnote{See 
http://supernova.lbl.gov/$\sim$nugent/deepsky.html.}.  The galaxy \hosts\ is
only weakly detected in most images.  We therefore forgo image
subtraction and perform photometry with a 2\farcs5 (radius)
aperture on all individual frames,
using the SDSS $i^{\prime}$ filter to calibrate the observed red bandpass
(an order-blocking filter with a cutoff blueward of $\lambda \approx
6100$\,\AA).  The resulting photometry is given in
Table~\ref{tab:deepsky}.

Taking the weighted mean of all the pre-outburst 
Deep Sky measurements of the potential host galaxy, 
we calculate $\langle i^{\prime} \rangle = 19.53 \pm 0.17$\,mag, 
in good agreement with the value from SDSS given the somewhat 
different passbands ($i^{\prime}_{\mathrm{SDSS}} = 1
9.65 \pm 0.04$\,mag).  If we assume a
constant flux equal to this value, we find $\chi^{2} = 
6.32$ (17 degrees of freedom, d.o.f.).  The observations are therefore 
consistent with a static flux level (null probability $0.991$).

We caution, however, that an outburst of comparable magnitude to that
observed from PTF10iya ($i^{\prime} = 20.24$ mag; \S\ref{sec:obs:p60})
may not have been detectable due to the small ratio of host-to-transient
flux and the low signal-to-noise ratio of most observations.  The
above limits therefore more directly limit brighter and/or redder
outbursts from \hosts.  

\input{tab1.tex}

\subsection{Palomar 60-inch photometry}
\label{sec:obs:p60}
Upon discovery of PTF10iya, the field was automatically inserted
into the queue of the robotic Palomar 60-inch telescope (P60;
\citealt{cfm+06}) for multi-colour follow-up observations.  Images were
processed using our custom real-time pipeline, and then subtracted
from reference frames obtained several months after the outburst using
HOTPANTS (Figure~\ref{fig:finder1}).  Subtracting late-time (after 
2010 July) P60 images from archival SDSS frames yielded no 
residual flux, confirming  that the transient was below our detection
threshold at this time.  A log of our P60 observations of PTF10iya, 
with later images stacked to increase depth, is provided in 
Table~\ref{tab:optuv}.

\begin{figure*}
  \includegraphics[width=18cm]{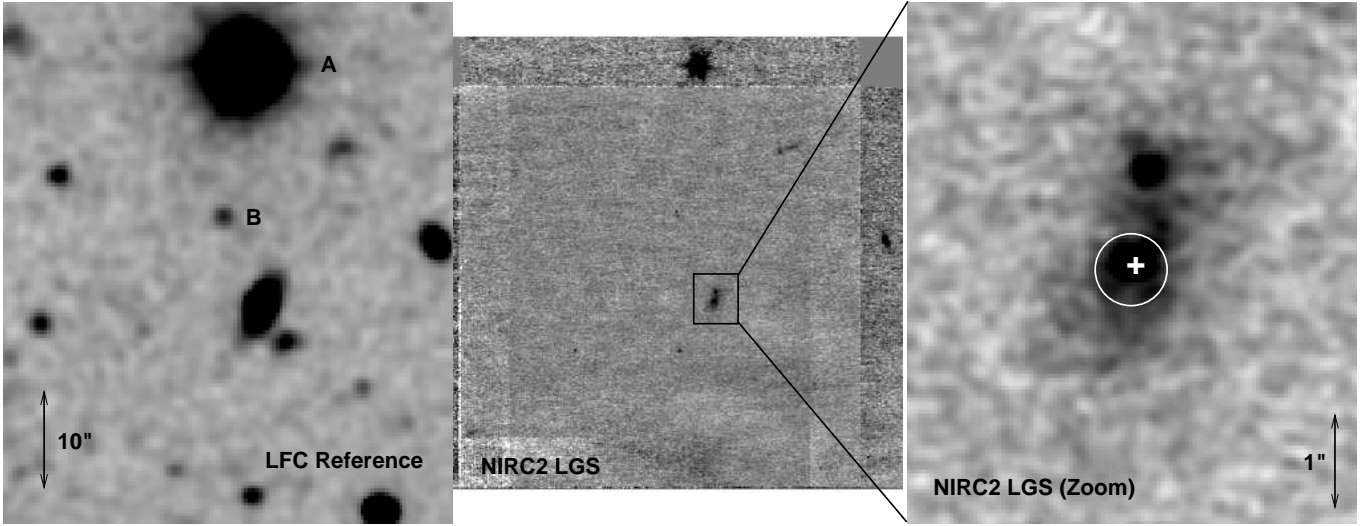}
  \caption{PTF10iya astrometry.  The left panel displays the Palomar LFC 
  $r^{\prime}$ reference imaging of the field of PTF10iya.  Due to the 
  relative sparseness of the field, the image contains only two point 
  sources in common with the NIRC2/LGS imaging (middle panel), 
  marked ``A'' and ``B.''  Using only these two sources, we calculate
  the 99.7\% containment radius for PTF10iya to be 350\,mas.  This
  localisation region is shown in a zoomed-in version of the NIRC2/LGS
  imaging in the right panel.  The position of the nucleus of \hosts, 
  marked with a plus sign, is consistent with the location of the 
  transient emission.  The point source to the NW is ruled out as a
  counterpart to PTF10iya at large confidence, and the lack of
  association is further reinforced by the absence of variability of this
  point source over the course of six months.  All images are oriented
  with north up and east to the left.}
\label{fig:finder2}
\end{figure*}

\subsection{\Swift\ UVOT/XRT}
\label{sec:obs:swift}
Motivated by the blue continuum (\S\ref{sec:obs:spec}) and the large
absolute magnitude of the outburst ($M_{R} \approx -21$\,mag;
\S\ref{sec:sed}), we triggered UV and X-ray target-of-opportunity 
observations of PTF10iya with the \Swift\ satellite \citep{gcg+04}.
The field was observed by the UV-Optical Telescope (UVOT;
\citealt{rkm+05}) and the X-ray Telescope (XRT; \citealt{bhn+05})
beginning at 17:53 on 2010 June 11.  A second set of images
  was obtained with \Swift\ on 2010 August 10, while a series of
  reference orbits observed from 2011 August 18 to 2011 
  September 7 were stacked to remove quiescent flux from the 
  potential host galaxy \hosts. 

For the UVOT $U$-band observations, we determined the count rate at the
location of PTF10iya using the techniques described by
\citet{ljf+06}, while for the $UVW1$, $UVM2$, and $UVW2$ data we used
the methods of \citet{pbp+08}.\footnote{We do not
  use the $V$-band and $B$-band data here, as the host plus transient were
  only marginally detected, and no \Swift\ reference frames were
  obtained in these filters.} 
After subtracting the pile-up corrected count rates
directly (see, e.g., \citealt{bhi+09}), transient emission at the location
of PTF10iya is clearly detected in all four blue filters in
  our initial epoch on 2010 June 11.  Comparing our 
  second epoch on 2010 August 10 with the late-time reference 
  images, we find that the UV flux has not varied over this interval.
  This would suggest that the transient emission has faded below the
  UVOT sensitivity limit several months after discovery.  A full
  listing of the host-subtracted UV photometry is provided in
  Table~\ref{tab:optuv}.   

We reduced the XRT data using the pipeline described by \citet{b07}.  
In the first epoch on 2010 June 11, an X-ray source is clearly
detected ($34.5 \pm 4.5$\,cts) at the location of PTF10iya (though with 
large astrometric uncertainty, $\sim 5$\arcsec), corresponding to a 
flux $F_{\rm X} = (7.5_{-2.5}^{+6.2}) \times 10^{-13}$\,erg\,cm$^{-2}$\,s$^{-1}$
(0.3--10\,keV).  Fitting a power-law model ($dN / 
dE \propto E^{-\Gamma}$) to the observed spectrum
results in a photon index of $\Gamma = 1.8_{-1.0}^{+1.2}$ ($\chi^{2} 
= 14.72$ for 17 d.o.f.).   No X-ray emission is detected at 
the location of PTF10iya in the second epoch obtained on 2010 
August 10, nor in the final reference epoch from 2011 
August -- September.  Assuming the same spectrum as in the 
first epoch, we derive a 3$\sigma$ flux limit (0.3--10\,keV) on
the quiescent flux level of $F_{\rm X} < 3 \times 
10^{-14}$\,erg\,cm$^{-2}$\,s$^{-1}$. 

\subsection{Keck laser guide-star adaptive optics}
\label{sec:obs:lgs}
On 2010 June 18, we observed the field of PTF10iya with
the Near-Infrared Camera 2 (NIRC2) mounted behind the
laser guide-star adaptive optics (LGS/AO) system on the Keck II telescope
\citep{wlb+06}.  Beginning at 8:10, we obtained a series of 
$K^{\prime}$-band exposures, each consisting of ten nondestructive 
readouts of either 10\,s or 20\,s, for a total time on source of 800\,s.
Images were reduced using standard IRAF\footnote{IRAF is 
distributed by the National Optical Astronomy Observatory, which is 
operated by the Association for Research in Astronomy, Inc., under 
cooperative agreement with the National Science Foundation.} 
routines, using a median combination of the (nonaligned) dithered 
images to correct for the sky background.  Prior to registration, 
we applied the distortion correction provided by the Keck 
Observatory\footnote{See
  http://www2.keck.hawaii.edu/inst/nirc2/forReDoc/post\_observing/dewarp.}.  
The resulting image (shown in the middle and right panels of 
Figure~\ref{fig:finder2}) was used for astrometric analysis 
(\S\ref{sec:astrometry}).  However, this procedure does not conserve 
flux, and so we created a separate coaddition that did not include 
the distortion correction for photometry.

A second series of LGS/AO images was obtained on 2010
December 2 with a total integration time of 750\,s using the identical 
instrumental setup.  Both \hosts\ and the offset point source 
(\S\ref{sec:astrometry}) are detected, with no evidence for 
variability in either source.

\subsection{Palomar 200-inch Large Format Camera}
\label{sec:obs:lfc}
We observed the field of PTF10iya with the Large Format Camera (LFC; 
\citealt{sms+00}) mounted on the 5-m Palomar Hale Telescope on 
2010 August 9.  A total of five dithered exposures (each 120\,s) 
were obtained in the $r^{\prime}$ filter beginning at 4:46.  After 
registering the individual images to a common reference, the frames 
were combined into a single coadd using the Swarp software
package\footnote{See http://www.astromatic.net/software/swarp.}.  The
resulting stacked image is shown in the left panel of Figure~\ref{fig:finder2}. 

\input{tab2.tex}

\subsection{Optical spectroscopy}
\label{sec:obs:spec}
We undertook a series of spectroscopic observations of PTF10iya with
the Intermediate dispersion Spectrograph and Imaging System 
(ISIS) mounted on the 4.2-m William Herschel Telescope 
(WHT; June 9) and the Low Resolution Imaging Spectrometer (LRIS;
\citealt{occ+95}) mounted on the 10-m Keck I telescope (June 8,
June 12, July 7).  For all spectra the slit was oriented at the
parallactic angle to minimise losses due to atmospheric dispersion
\citep{f82}.  Details of the observational setup for each respective
spectrum are provided in Table~\ref{tab:spec}.

All spectra were reduced using standard routines (see, e.g.,
\citealt{esn+08} for details).  Spectra were 
extracted optimally \citep{h86} within the IRAF environment.   A 
dispersion solution was computed using afternoon calibration spectra 
of arc lamps, and then adjusted for each individual exposure using
night-sky lines.  Telluric atmospheric absorption features were 
removed using the continuum from spectrophotometric standard stars
\citep{wh88,mfh+00}.  Finally, a sensitivity function was applied
using observations of spectrophotometric standards at a 
comparable airmass.  The red and blue arms were rebinned to a common
dispersion and then joined across the dichroic.

\begin{figure}
  \includegraphics[width=9cm]{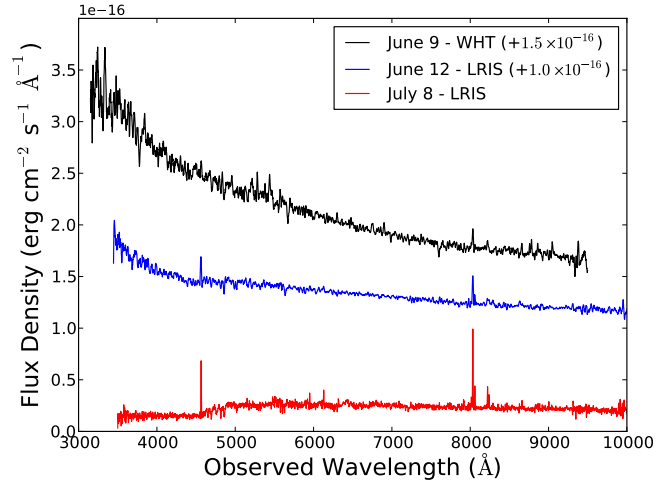}
  \caption{Spectroscopic observations of PTF10iya/\hosts.  The spectra from
  June 9 (WHT) and June 12 (LRIS) have been offset for clarity.
  Our spectrum from June 8 obtained with LRIS is not
  plotted, as it is nearly identical to the WHT spectrum.}
\label{fig:spec}
\end{figure}

To account for slit losses, we have adjusted the flux calibration for
all four spectra using broadband photometry from our (unsubtracted)
P60 images at comparable epochs. Three of the resulting 
spectra are shown in Figure~\ref{fig:spec}.

\subsection{Radio Observations}
\label{sec:obs:radio}
We observed the location of PTF10iya with the 
  NRAO\footnote{The National Radio Astronomy
  Observatory (NRAO) is a facility of the National Science Foundation
  operated under cooperative agreement by Associated Universities,
  Inc.} Expanded Very Large Array (EVLA) for 1 hour on 2011 April 29.3 UT.
  Observations with a total bandwidth of 256 MHz were obtained
  at centre frequencies of 
  1.3 and 8.4 GHz. A compact source (J1426+3625) near PTF10iya was 
  observed every 4--6 min for accurate phase calibration, while 
  3C\,286 was observed at the end of the run for the bandpass and 
  flux-density calibration.  The data were reduced and imaged using the 
  Astronomical Image Processing System (AIPS) software package.

There was no emission detected at the position of PTF10iya  
  to 3$\sigma$ limits of 0.13\,mJy and 0.07\,mJy at 1.3 and 8.4 GHz, 
  respectively. At the distance of the host galaxy (\S\ref{sec:host}), 
  this corresponds to spectral luminosities of $<$1.4$\times 
  10^{29}$\,erg\,s$^{-1}$\,Hz$^{-1}$ (1.3 GHz) and 
  $<$ 7.8$\times 10^{28}$\,erg\,s$^{-1}$\,Hz$^{-1}$ (8.4 GHz).

\subsection{Additional archival data}
\label{sec:obs:archives}
As discussed in \S\ref{sec:obs:p48}, PTF10iya is spatially
coincident (see \S\ref{sec:astrometry} for a detailed discussion of
astrometry) with the extended source \hosts.  
The photometry (model magnitudes) provided by SDSS 
implies a modestly red galaxy ($g^{\prime} - r^{\prime} = 0.72$\,mag).  

The location of PTF10iya was observed with the Very Large Array (VLA) 
at 1.4\,GHz in 1995 December as part of the Faint Images of the Radio Sky at
Twenty centimeters (FIRST; \citealt{bwh95}).  No
source is detected at this location to a 3$\sigma$ limiting flux
density of 0.39\,mJy.
There is no source consistent with the location of PTF10iya in
either the \fermi\ Large Area Telescope (LAT; 100\,MeV -- 100\,GeV) 
1-year Point Source Catalog \citep{aaa+10e} nor the {\it ROSAT} (0.1 -- 
2.4\,keV) All-Sky Survey Bright Source Catalog \citep{vab+99}.     

\section{Analysis}
\label{sec:analysis}

\subsection{Astrometry}
\label{sec:astrometry}
We first determine if there is any discernible offset between the observed
transient emission and the nucleus of the galaxy \hosts, as this
greatly constrains viable models for the outburst.  Because of the
significantly smaller pixel scale (0\farcs38, compared with
1\farcs0 pixel for the P48 camera), we initially consider the 
P60 imaging.  

Starting with the $g^{\prime}$ data (where the transient emission is
detected most strongly), we compute an astrometric solution for the 
reference image using $20$ point sources in SDSS.  The 1$\sigma$ 
uncertainty associated with the absolute astrometric calibration is 
90\,mas in each coordinate.  Using the centroiding method from 
SExtractor\footnote{See http://www.astromatic.net/software/sextractor.},
we measure a position for \hosts\ in our reference images 
of $\alpha = 14^{\mathrm{h}} 38^{\mathrm{m}} 40.988^{\mathrm{s}}$, $\delta =
+37^{\circ} 39\arcmin 33\farcs43$ (J2000.0), consistent within the
stated uncertainties with the SDSS measurement.

Aligning our P60 $g^{\prime}$ image from 2010 June 7 with the coadded
P60 reference image using 15 common point sources, we measure a 
dispersion of 0.20\,pixel (80\,mas) in each direction.  On
the subtracted image, we measure a position for the transient emission
of $\alpha = 14^{\mathrm{h}} 38^{\mathrm{m}} 40.983^{\mathrm{s}}$, $\delta =
+37^{\circ} 39\arcmin 33\farcs46$, corresponding to an offset from
\hosts\ of $(+0.09, -0.14)$ pixels.  Repeating the process for the $r^{\prime}$
and $i^{\prime}$ filters yields similar results.  Thus, based solely
on the P60 observations, we conclude that PTF10iya occurred within 
120\,mas (radius) of the nucleus of \hosts\ (68\% confidence interval).  
At $z = 0.224$ (\S\ref{sec:host}),
this corresponds to a projected distance of $d < 420$\,pc.

The significantly improved angular resolution provided by the LGS/AO 
imaging system with NIRC2 could provide even tighter constraints
on any offset between PTF10iya and \hosts.  It is clear from the
LGS image (Figure~\ref{fig:finder2}, right panel) that a point source
is detected offset from the nucleus of \hosts: using the NIRC2 (wide 
camera) pixel scale (0\farcs039686), we measure a radial offset of
970\,mas (corresponding to a cardinal offset of 140\,mas W,
960\,mas N at the specified position angle of 0$^{\circ}$). 

Aligning the NIRC2 data from 2010 June 18 with optical imaging of the 
field is complicated by the lack of sources common to both frames.  
Even in our deepest LFC stack, we find only two point 
sources\footnote{The requirement that the common sources be unresolved
  by default precludes the usage of \hosts.} present in
both images (Figure~\ref{fig:finder2}, sources ``A'' and ``B'').  
We therefore allow for only a translational offset (i.e., $\Delta x$, 
$\Delta y$) between the two images.  In other words, we have fixed 
the relative scale and rotation (based on the known NIRC2 pixel 
scale and position angle), and calculate the average shift between 
the two point sources.  The resulting uncertainty associated with 
this procedure, calculated as the average offset between the position 
of sources A and B in the LFC and NIRC2 images, is 50\,mas in
each coordinate.

Using the resulting astrometry,
we measure a position for the point source of $\alpha =
14^{\mathrm{h}} 38^{\mathrm{m}} 40.968^{\mathrm{s}}$, $\delta =
+37^{\circ} 39\arcmin 34\farcs43$.    Thus, the point source is offset
from the transient location by 990\,mas in radius.  Even if we assume
a relatively conservative uncertainty of 100\,mas in each coordinate
(double the derived value) to account for the lack of common sources
for registration, we still conclude that the point source is offset from
PTF10iya with overwhelming confidence.

On the other hand, in the NIRC2 image we measure a position for \hosts\
of $\alpha = 14^{\mathrm{h}} 38^{\mathrm{m}} 40.980^{\mathrm{s}}$, 
$\delta = +37^{\circ} 39\arcmin 33\farcs47$.  This falls only
40\,mas from the location of PTF10iya, fully consistent within the
errors.  This is clearly illustrated in the right panel of
Figure~\ref{fig:finder2}, where the centroid of \hosts\ (indicated with
a plus sign) falls within the 99.7\% confidence (350\,mas 
adopting our conservative estimate) localisation of PTF10iya.  
The point source, however, is well outside this position.  

Two additional lines of evidence further argue against an association
between the offset NIR point source and the optical transient
PTF10iya.  Using the $K$-band magnitude for the nearby bright star 
2MASS\,J143841.12+3739571 (Source A in Figure~\ref{fig:finder2})
from the Two Micron All-Sky Survey Point Source Catalog \citep{scs+06}
as a reference, we measure a 
magnitude of $K = 19.61 \pm 0.08$ (Vega) for the offset point source
on 2010 June 18.  Within uncertainties, this value remains constant in
our next epoch of $K$-band imaging nearly 6 months later.  In addition, 
in terms of flux density ($f_{\nu}$), the point source is actually
brighter than the contemporaneous optical limits from P60.
Given that the observed spectrum of PTF10iya was extremely blue
(\S\ref{sec:sed}), it seems even more unlikely that this point source
is associated with PTF10iya.

As the offset NIR point source is undetected in even our deepest
(P200) optical imaging of the field, the object could either be a
foreground cool dwarf star, or an unresolved overdensity in
\hosts\footnote{While M dwarfs are known
  to undergo dramatic, blue outbursts and can sometimes be confused
  for extragalactic transients (e.g., \citealt{kr06}), the timescale
  of PTF\,10iya is orders of magnitude longer than any known such
  outburst.}.  We shall assume it is unrelated to PTF10iya for the
remainder of this work.  

To summarise, we conclude that the location of the transient
PTF10iya is consistent with the nucleus of the galaxy \hosts.  Our
99.7\% confidence localisation, with a radius of 350\,mas, corresponds
to a projected distance of 1.2\,kpc at $z = 0.224$ (the 68\% 
confidence radius of 150 mas corresponds to a projected distance of
only 540\,pc).  Given the astrometric alignment, together with the
detection of absorption features in the transient spectra 
($\S$\ref{sec:sed}), we shall assume for the remainder of
this work that PTF10iya is associated with \hosts.

\begin{figure}
  \includegraphics[width=9cm]{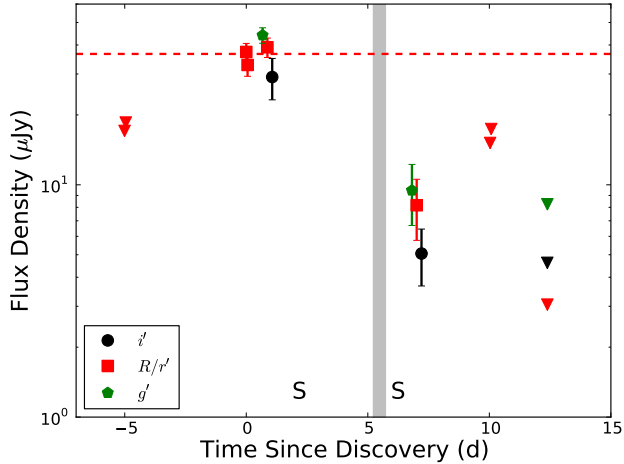}
  \caption{Observed optical light curve of PTF10iya, around the time
    of outburst.  The time is referenced to the P48 discovery on Jun
    6.302.  Inverted triangles represent 3$\sigma$ upper limits.
    The shaded grey vertical bar indicates the epoch of our
    \Swift\ X-ray and UV observations.  The horizontal dashed line
    indicates the quiescent ($r^{\prime}$) magnitude of the host
    galaxy \hosts.  The times of optical
    spectroscopic observations are marked with an ``S.''  No
    correction has been applied for extinction within the host galaxy
    (\S\ref{sec:host}).  Note that the $i^{\prime}$ and $g^{\prime}$
    data have been offset slightly in the horizontal direction for
    plotting purposes.}
\label{fig:lcurve}
\end{figure}

%
\subsection{Light curve, spectral energy distribution, 
                    and energetics}
\label{sec:sed}
The optical light curve of PTF10iya (i.e., after subtracting 
host-galaxy contamination) is shown in Figure~\ref{fig:lcurve}.  Given the
nondetection with P48 on June 1, we can place a limit on the rise
time of $\tau_{\mathrm{r}} < 5$\,d ($dm/dt >
0.15$\,mag\,d$^{-1}$).  The $R/r^{\prime}$ flux remains
roughly constant for at least one day, after which it declines quite
steeply at a rate of $\gtrsim 0.3$\,mag\,d$^{-1}$, or an $e$-folding
timescale of $\tau_{\mathrm{d}} \approx 4$\,d.  A similar decay is
seen in both the $g^{\prime}$ and $i^{\prime}$ filters.

Our detections of PTF10iya are largely clustered into two distinct
epochs: June 6--9 (P48 and P60 optical photometry, WHT and LRIS optical
spectroscopy), and June 11--13 (P60 optical photometry, \Swift\ UV and
X-ray photometry, and LRIS spectroscopy).  With only limited
constraints on the light-curve evolution, we assume that the flux 
in all bandpasses is approximately constant over the
course of each of the two epochs, and proceed to study the 
broadband spectral energy distribution (SED).

In order to do so, we  must first (1) remove the contribution of the
host galaxy \hosts\ from our spectroscopic observations, and 
(2) deredden the (subtracted) spectra and photometry to account for 
extinction within the host galaxy.  As described in
$\S$\ref{sec:obs:spec}, the absolute flux scale for each spectrum has
been calculated using contemporaneous P60 photometry of the host plus
transient (i.e., before the image-subtraction process).  To remove
the host-galaxy contribution from our June spectra of PTF\,10iya, we thus
subtracted the observed LRIS spectrum from July 7 (which is
dominated by host light).

\begin{figure*}  
  \begin{minipage}[b]{0.45\linewidth}
  \centering
  \includegraphics[width=9cm]{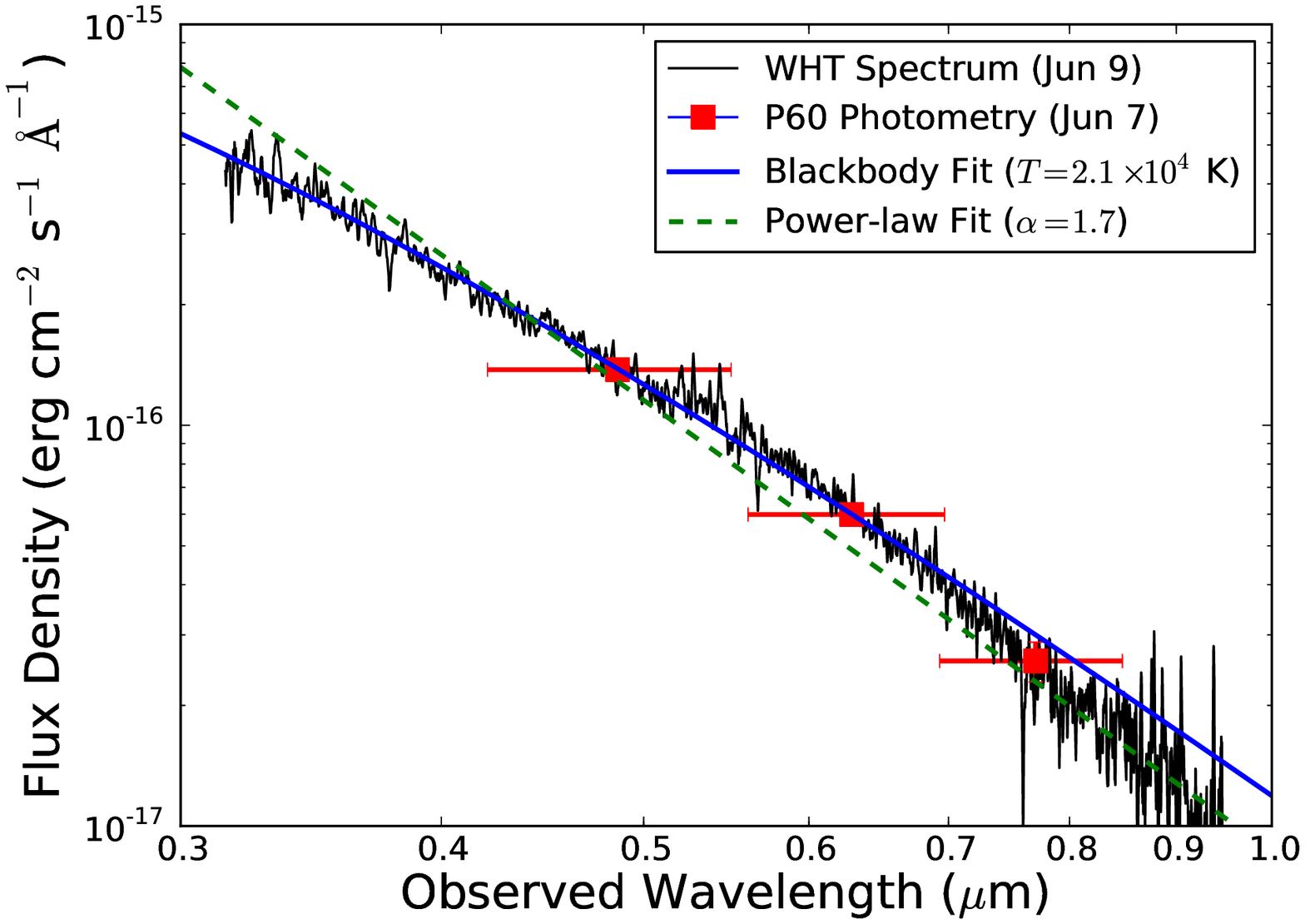}
  \end{minipage}
  \hspace{0.5cm}
  \begin{minipage}[b]{0.45\linewidth}
  \centering
  \includegraphics[width=9cm]{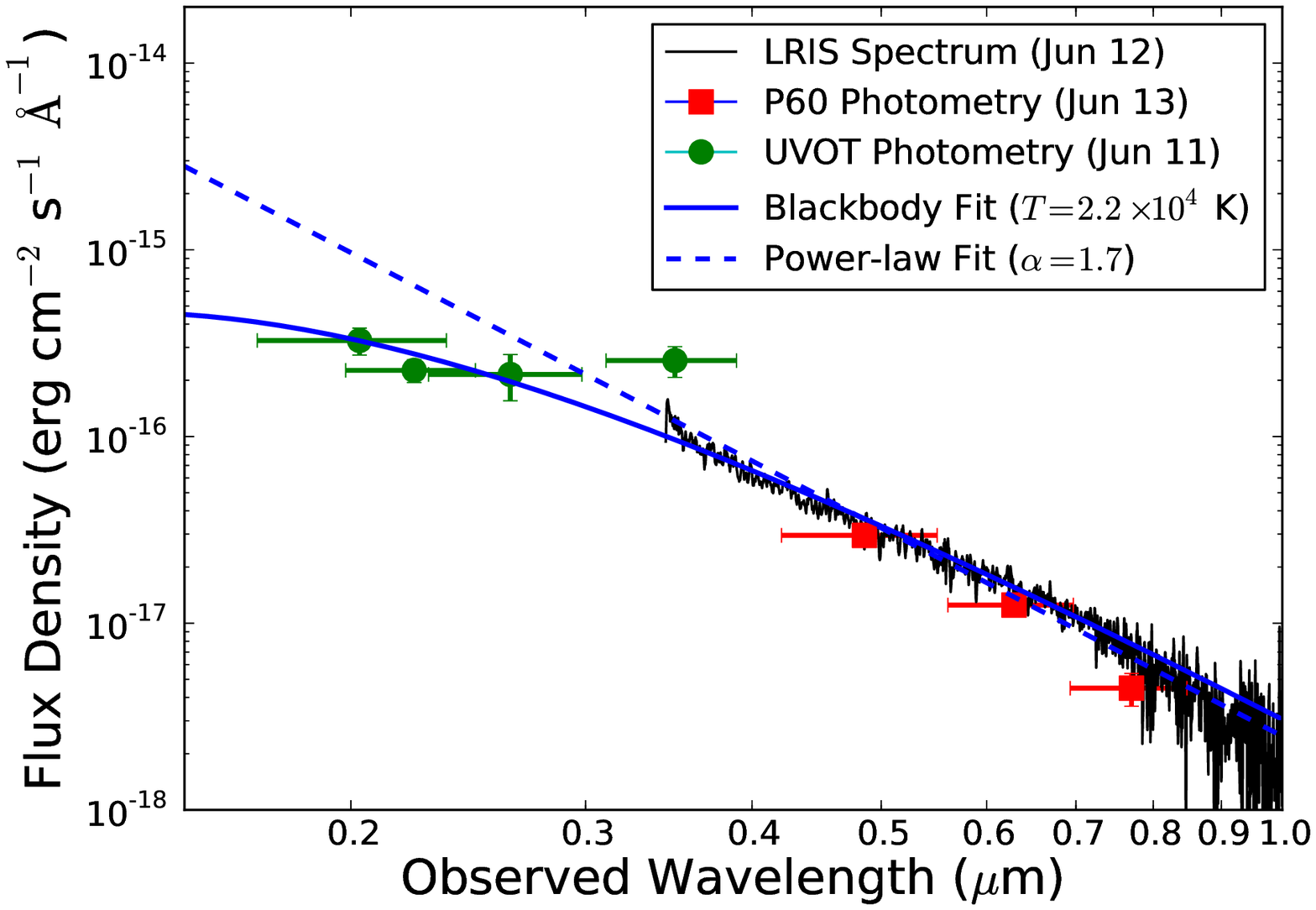}
  \end{minipage}
  \caption{Host-galaxy subtracted and extinction-corrected SED of 
    PTF10iya.  \textit{Left panel:} The June 9 WHT spectrum, with P60
    (host-subtracted) photometry overlaid.  A blackbody fit results in a
    temperature of $T_{1} = 2.1 \times 10^{4}$\,K.  The power-law model
    (dashed line) does not reproduce the observed curvature in the spectrum 
    as well as a Planck function.  \textit{Right panel:} The June 12 LRIS
    spectrum.  Also overlaid are host-subtracted optical
    photometry from P60 and UV photometry from \Swift/UVOT.
    The best-fitting blackbody function, with $T_{2} = 2.2
    \times 10^{4}$\,K, is plotted as a solid line.  The fit quality is
  not as high as the previous epoch, but the Planck function better
  reproduces the turnover in the SED blueward of 3000\,\AA\ than a
  simple power-law fit (dashed line).}
\label{fig:subtracted}
\end{figure*}

While the initial spectra contained both emission and
absorption lines consistent with the redshift of \hosts, the resulting
subtracted spectra are largely featureless with a relatively strong
blue continuum (Figure~\ref{fig:subtracted}).  All residual features
are narrow and correspond to nebular emission or stellar absorption
lines seen in the late-time host spectrum, and thus could result from small
mismatches in resolution, slit orientation, or wavelength solution.

To account for host-galaxy extinction, we use the reddening law
inferred for starburst galaxies from \citet{c01} with
$E(B-V)_{\mathrm{gas}} = 0.40$ mag, which we derive from the observed
Balmer decrement ($\S$\ref{sec:host}).  We discuss the effect
of uncertainties in the inferred extinction on our results in what
follows.

In Figure~\ref{fig:subtracted} we plot the derived host- and
extinction-corrected UV/optical SED of PTF10iya at both epochs.  
The broadband photometry and spectra agree reasonably well and 
suggest that any systematic uncertainties introduced in this process are 
relatively modest.

We consider several continuum models in attempting to fit the observed
UV/optical SEDs.  A power-law model ($f_{\lambda} \propto
\lambda^{-\alpha}$) fails to reproduce the observed spectral curvature
(particularly for the June 6--9 data; left panel of
Figure~\ref{fig:subtracted}), and significantly overpredicts the UV
flux.  A blackbody spectrum, however, provides a reasonable
description of the data at both epochs.  The best-fitting blackbody models
are plotted in Figure~\ref{fig:subtracted}, while the derived fit
parameters are displayed in Table~\ref{tab:bbfits}.

\input{tab3.tex}

Two sources of uncertainty could potentially introduce large
systematic errors into these results.  First, improper flux
calibration (particularly in the near-UV, where atmospheric absorption
can be highly variable) or host-galaxy subtraction could bias the
derived blackbody parameters.  Our nearly simultaneous LRIS (June
8.43) and WHT (June 9.10) spectra allow us to test this to some
extent, as the spectrum is unlikely to have evolved dramatically 
over this period.  We find similar fit parameters for
both spectra, although the spread is significantly larger than the
formal error from a single fit.  We therefore adopt the difference 
in the derived fit parameters between two spectra as our 
estimated uncertainty for the June 6--9 epoch.  A comparable fractional
uncertainty is applied to the data from June 11--13, where only a
single spectrum is available.

Second, because the optical bandpass falls on the Rayleigh-Jeans tail,
small changes to the extinction correction (either the magnitude or
the dependence with wavelength) could significantly impact the derived
fit parameters.  Most importantly, minor changes in the UV flux and 
derived temperature could dramatically affect the derived blackbody
luminosity, as $L_{\mathrm{BB}} \propto T^{4}$.  We therefore repeat
the above analysis, fitting blackbody spectra to the uncorrected SEDs
[i.e., $E(B-V) = 0$ mag], and treat these results as lower limits on the
blackbody luminosity and temperature.  The results of this analysis
are shown in Table~\ref{tab:bbfits}.

Independent of the details of the extinction correction, we conclude
that the UV/optical SED of PTF10iya is reasonably well fit by a blackbody
spectrum with temperatures of order a few times $10^{4}$\,K.  The
temperature does not vary dramatically over the course of our
observations (June 6--13), though the luminosity does decrease by a
factor of 3--4.  Likewise, the blackbody radius appears to
decrease during this period, although this result is not particularly
significant.

\begin{figure}
  \includegraphics[width=9cm]{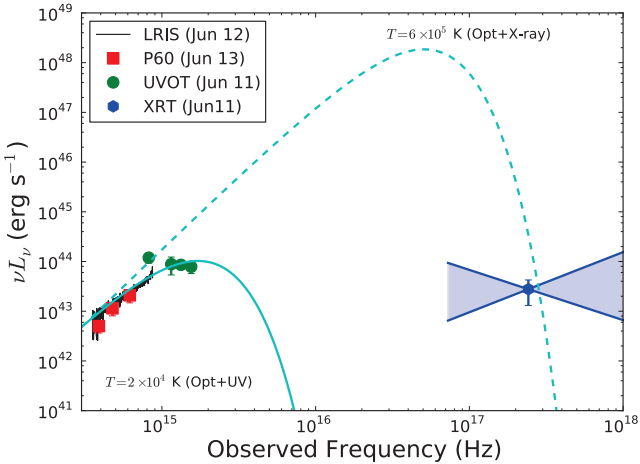}
  \caption{Broadband SED of PTF10iya.  We have plotted the LRIS
    spectrum from June 12, P60 photometry from June 13, and UVOT
    and XRT photometry from June 11.  A
    blackbody fit to incorporate both the optical and X-ray data
    requires $T \approx 6 \times 10^{5}$\,K, but is inconsistent with
    the observed X-ray spectrum (the range of allowed power-law indices
    is indicated by the shaded blue region) and severely overpredicts
    the UV flux.  On the other hand, a blackbody fitting both the UV
    and optical flux ($T = 2 \times 10^{4}$\,K) cannot account
    for the bright X-ray emission. }
\label{fig:sed}
\end{figure}

In Figure~\ref{fig:sed}, we plot the broadband (optical, UV, and
X-ray) extinction-corrected SED of PTF10iya, this time in terms of
$\nu L_{\nu}$.  The best-fitting UV/optical 
blackbody, with $T = 2 \times 10^{4}$\,K, significantly 
underpredicts the X-ray flux.  While it is possible to fit the 
optical and X-ray data with a single blackbody with $T \approx 6 
\times 10^{5}$\,K (dashed cyan line), the fit is incompatible with the 
range of possible X-ray spectral indices (blue shaded region), and
severely overpredicts the UV flux.  We conclude that the X-ray emission 
is produced by a physical process distinct from that of the observed
UV/optical emission. 

In terms of $\nu L_{\nu}$, PTF10iya emitted a comparable amount of energy
in the X-rays as in the UV/optical bandpass. Over the range 0.3--10\,keV, 
we find $L_{\rm X} = 1 \times 10^{44}$\,erg\,s$^{-1}$, about
1--3 times the UV/optical blackbody luminosity at this time
(depending on the extinction correction).  To estimate the total
energy radiated, we multiply the mean blackbody luminosity by an
approximate duration of 10\,d, and find $E_{\mathrm{rad}} \approx
3 \times 10^{50}$\,erg.  

\begin{figure*}
  \includegraphics[width=18cm]{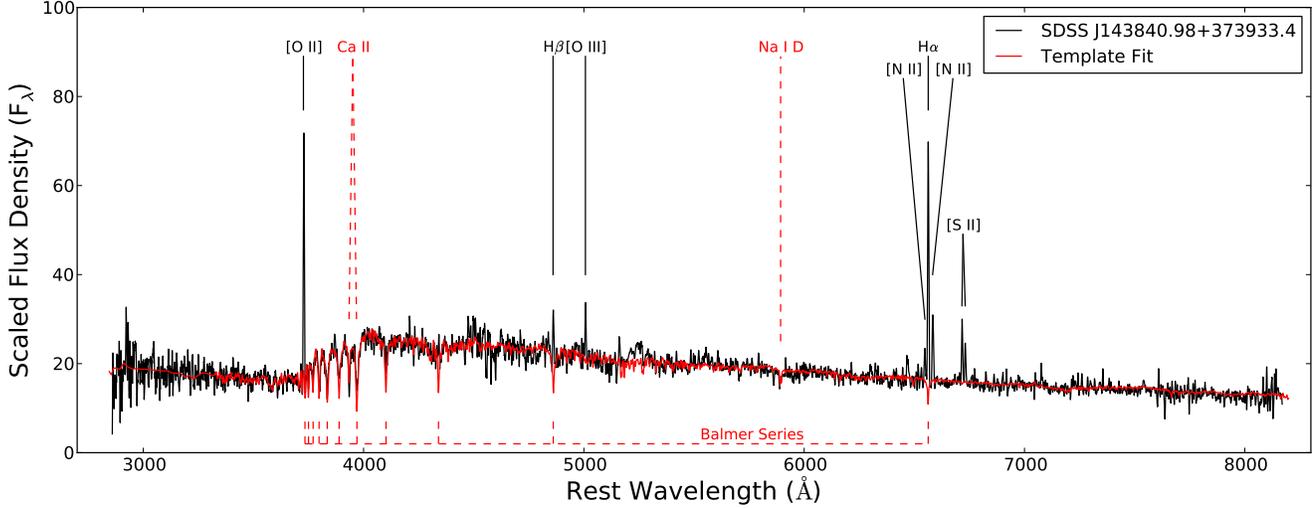}
  \caption{Deredshifted ($z=0.22405$) and extinction-corrected 
    [$E(B-V)_{\mathrm{gas}} = 0.40$ mag] host-galaxy spectrum 
    (black) and template fit (red).
    Following \citet{thk+04}, we fit a linear combination of 39
    stellar synthesis templates from \citet{bc03} in order to remove
    the stellar continuum from the spectrum of \hosts.  The resulting
    emission-line intensity ratios (Figure~\ref{fig:bpt}) were then 
    calculated using the subtracted spectrum.}
\label{fig:host}
\end{figure*}

%
\subsection{Host galaxy}
\label{sec:host}
The July 8 spectrum of \hosts, after dereddening (see below), is
plotted in Figure~\ref{fig:host}.
The spectrum is clearly a composite, containing both absorption
features from an underlying stellar population and a series of 
narrow nebular emission lines.
In particular, we detect Ca\,II H\&K, as well as H$\delta$,
H$\epsilon$, H$\zeta$, H$\eta$, H$\theta$, H$\iota$, and
H$\kappa$, all resolved in absorption.
Likewise, typical galaxy emission features,
including [O\,II] $\lambda$3727, H$\beta$, [O\,III]
$\lambda$5007, [N\,II] $\lambda$6548,
H$\alpha$, [N\,II] $\lambda$6584, and [S\,II] $\lambda
\lambda$6716, 6731, are all clearly detected from \hosts.  Using
the strongest unblended emission lines, we measure a redshift of
$z = 0.22405 \pm 0.00006$.  All of the emission lines are unresolved 
by the red arm of our LRIS spectrum, with a full width at
half-maximum intensity (FWHM) of $\sim 6.5$\,\AA.  

We can estimate the extinction along the line of sight to the 
source using the observed intensity ratios of host-galaxy Balmer 
emission lines.
Fitting a Gaussian profile to the lines, we find that
$(L_{\mathrm{H}\alpha} /
L_{\mathrm{H}\beta})_{\mathrm{obs}} = 4.4 \pm 0.4$.
Assuming Case B recombination \citep{o89} and the relation from
\citet{c01}, we find $E(B-V)_{\mathrm{gas}} = 0.40 \pm 0.08$ mag.

The fundamental issue we wish to resolve is the origin of these
emission lines.  The atoms could be ionised by the hard power-law spectrum
generated by gas accretion onto a central SMBH
(i.e., an AGN), UV photons from young, massive O and B stars (i.e.,
star formation), or as part of a phenomenon known as a Low Ionisation
Nuclear Emission-line Region (LINER; \citealt{h80}), which are likely
related to AGNs, possibly resulting from changes to the geometry of the
disc at low accretion levels (e.g., \citealt{h08}).  

Typically this is done via a diagnostic diagram
\citep{bpt81,vo87} which compares various emission-line intensity ratios.
In order to extract the line flux due to the
central source, however, we must first remove any contributions from
the underlying stellar population (in particular, Balmer-series
absorption; e.g., \citealt{hfs97}).  Following \citet{thk+04} and
\citet{kht+03}, we fit a version of the dereddened host spectrum with the
emission lines removed to a linear combination of a series of 39
template spectra constructed from the stellar synthesis models of
\citet{bc03}.  The resulting least-squares fit (restricting the
weighting coefficients to be positive) is shown in red in
Figure~\ref{fig:host}, and does a reasonable job reproducing the
observed host spectrum across the entire bandpass.

\begin{figure*}
  \begin{minipage}[b]{0.30\linewidth}
  \centering
  \includegraphics[width=6cm]{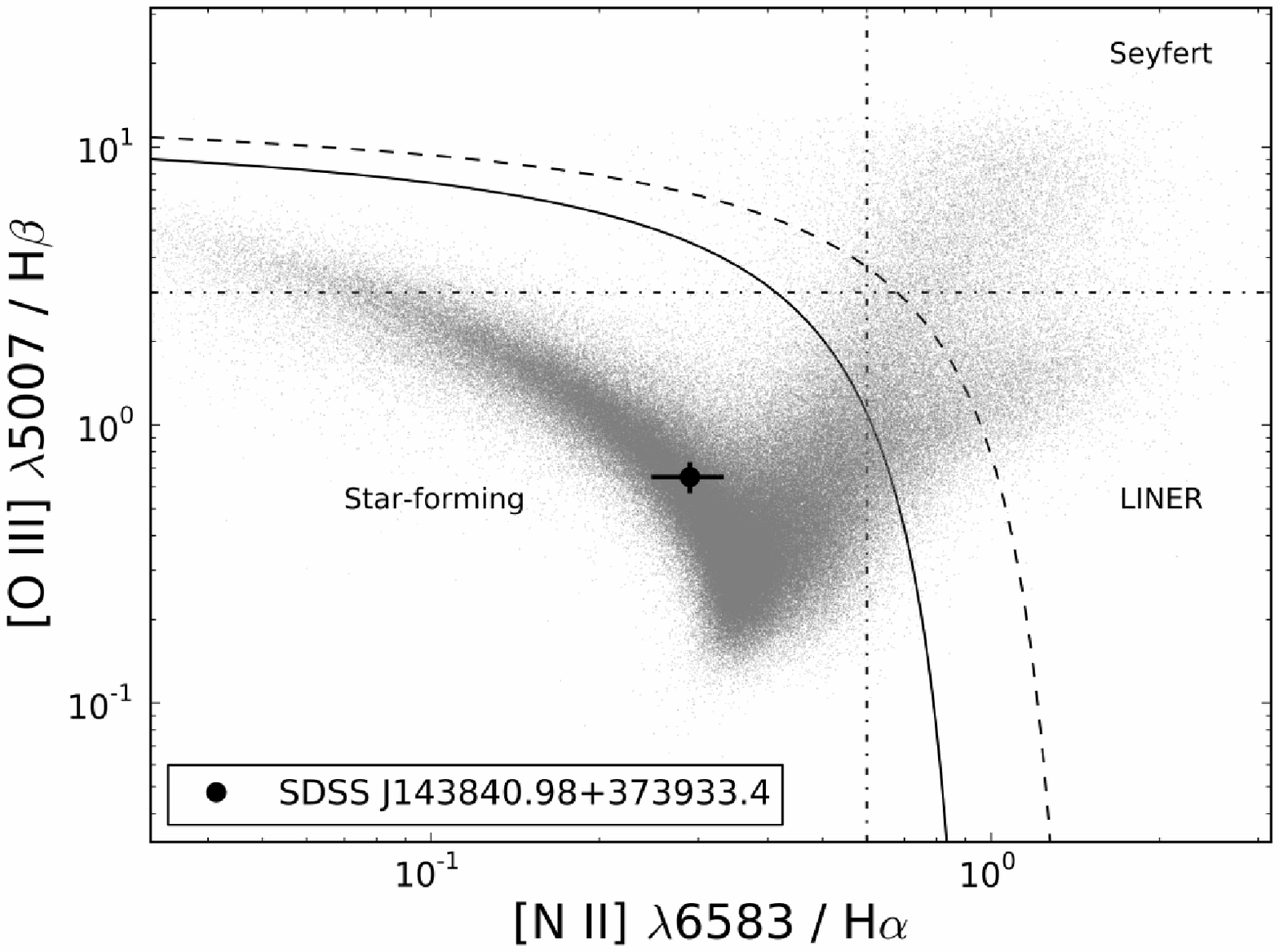}
  \end{minipage}
  \hspace{0.25cm}
  \begin{minipage}[b]{0.30\linewidth}
  \centering
  \includegraphics[width=6cm]{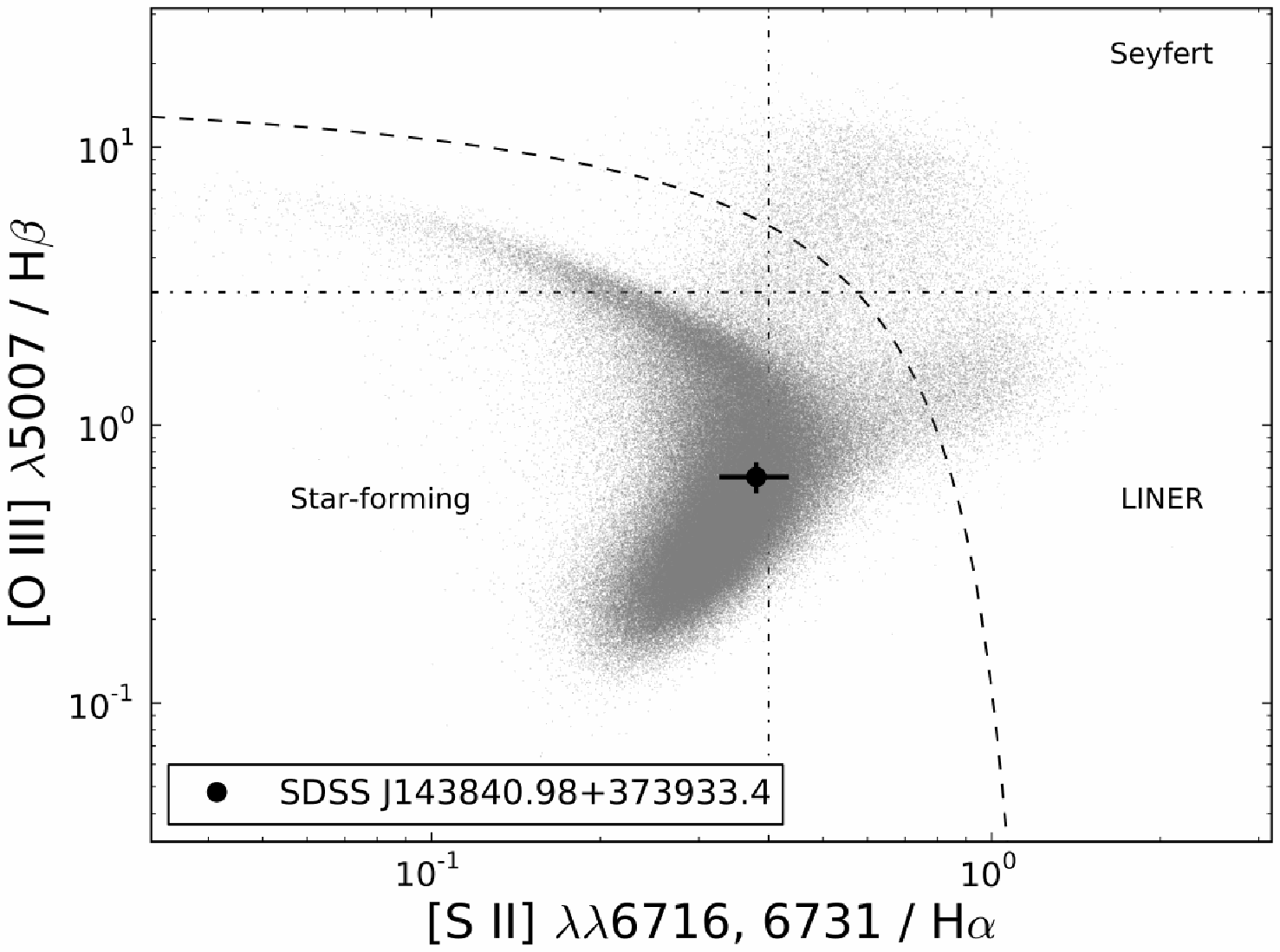}
  \end{minipage}
  \hspace{0.25cm}
  \begin{minipage}[b]{0.30\linewidth}
  \centering
  \includegraphics[width=6cm]{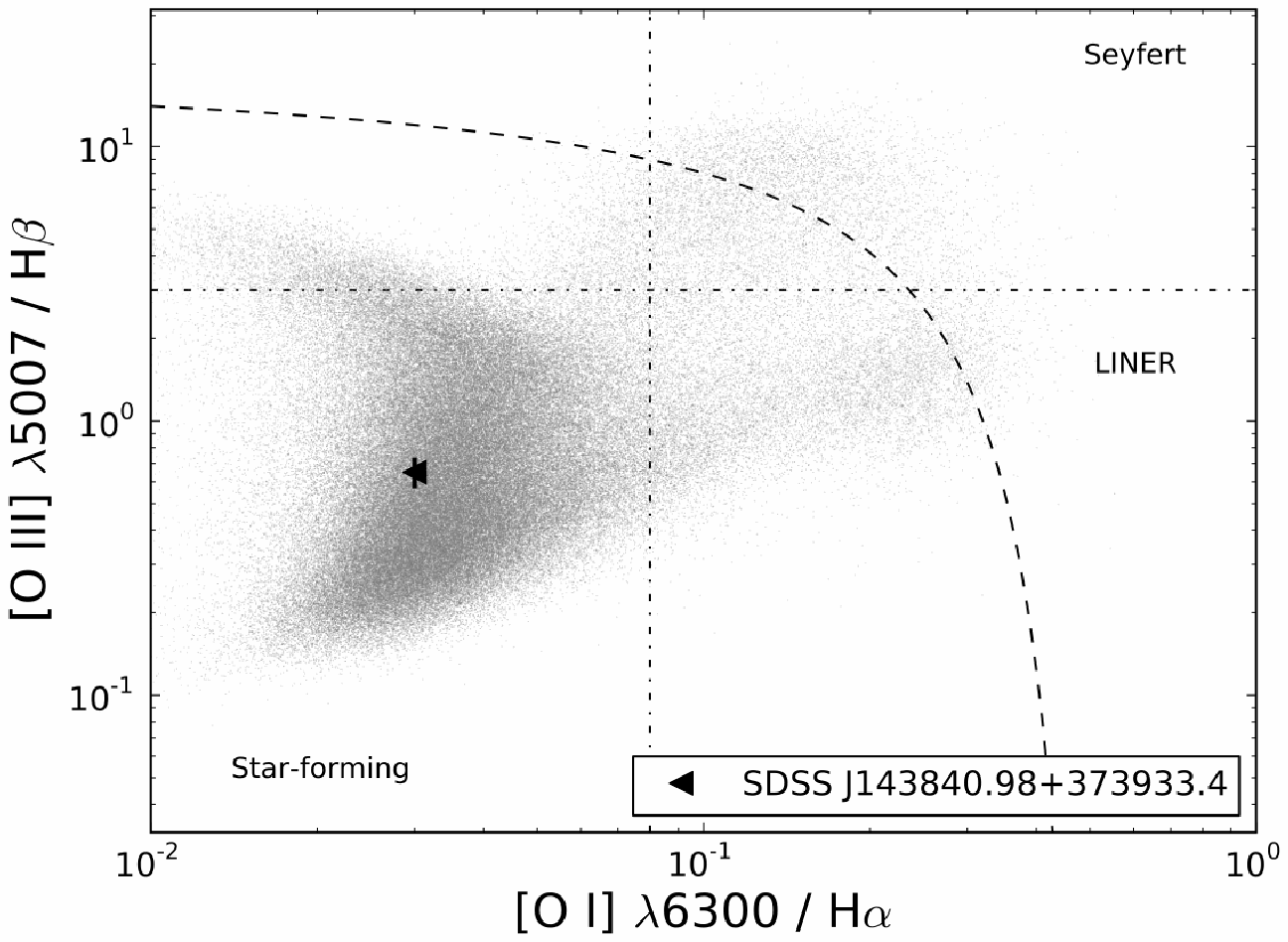}
  \end{minipage}
  \caption{Diagnostic emission-line diagrams for \hosts.  The empirical
  dividing line between star-forming and active galaxies from
  \citet{kht+03} is shown as the solid line in the left panel,
  while analogous dividing lines from \citet{hfs97} are indicated with
  dashed-dotted lines.  The theoretical dividing line from
  \citet{kd02} is plotted as a dashed line.  Analogous measurements
  for SDSS galaxy from the MPA-JHU value-added catalog are shown as
  gray dots.}
\label{fig:bpt}
\end{figure*}

After subtracting the template to remove stellar contamination, we
measure the following diagnostic ratios: $L_{\left[ \mathrm{O\,III} \right]
    \lambda 5009} / L_{\mathrm{H}\beta} = 0.65 \pm 0.07$, 
    $L_{\left[ \mathrm{N\,II} \right] \lambda 6583} / L_{\mathrm{H}\alpha} =
    0.29 \pm 0.04$, $L_{\left[ \mathrm{S\,II} \right] \lambda\lambda 6716, 6731} /
    L_{\mathrm{H}\alpha} = 0.38 \pm 0.05$, and $L_{\left[ \mathrm{O\,I} \right]
        \lambda 6300} / L_{\mathrm{H}\alpha} < 0.03$.
The resulting diagnostic diagrams, including
division lines between star-forming galaxies, AGNs, and LINERs from
\citet{kht+03}, \citet{kd02}, and \citet{hfs97}, are shown in Figure~\ref{fig:bpt}.
Also plotted are analogous measurements for a large number of
galaxy spectra from the MPA/JHU value-added SDSS catalog\footnote{See
  http://www.mpa-garching.mpg.de/SDSS.}.

In all three plots, \hosts\ falls firmly within the locus of
star-forming galaxies from SDSS.  There may be a slight
excess of [S\,II] $\lambda\lambda$6716, 6731 relative to
H$\alpha$, as the host lies directly on the dividing line between 
star-forming galaxies and LINERs from \citet{hfs97}.  But even on
this plot, \hosts\ appears to be more consistent with the star-forming
galaxies from SDSS than the LINER branch, and does not reach the
theoretical limit from \citet{kd02}.  The simplest interpretation, then, is that
the narrow emission lines in \hosts\ result from photoionisation by the
UV photons copiously produced by young, massive stars.  

The observed morphology of \hosts\ further reinforces the notion that
this galaxy either is or recently was actively forming new stars.
Using the \texttt{GALFIT} software \citep{phi+02}, we fit the
$r^{\prime}$ LFC images of \hosts\ to a S\'{e}rsic profile
\citep{s67}, allowing
the concentration parameter ($n$) and the effective radius ($r_{e}$)
to vary freely.  We find that a good fit (both in terms of $\chi^{2}$ per
degree of freedom and by visual inspection) is achieved when $n \approx 0.65$ and
$r_{e} \approx 1\farcs7$.  Analogous fits to the SDSS $g^{\prime}$ and
$r^{\prime}$ images yield similar results.  

The relatively small concentration number suggests that the emission from
\hosts\ is dominated by a disc-like profile; for bulge-dominated
systems, the radial profile is typically well-fit by a de Vaucouleurs
function ($n = 4$).  Adding a bulge profile ($n = 4$)
significantly reduces the fit quality.  By varying the
fit parameters and examining the residuals, we estimate that the ratio of
the bulge to total luminosity ($L_{\mathrm{B}} / L_{\mathrm{T}}$) is
$\lesssim 30$\%, a typical value for spiral galaxies.

With this (approximate) deconvolution, we can constrain the mass of
the central SMBH using the observed bulge luminosity vs. black hole
mass relation \citep{mtr98}.  Integrating our (dereddened)
spectrum of \hosts\ over the rest-frame $V$ band, we find $M_{V} =
-20.75 \pm 0.15$\,mag ($M_{V} = -19.82 \pm 0.17$\,mag 
neglecting any host extinction).  Applying the bulge-to-total 
luminosity correction from above ($L_{\mathrm{B}} / L_{\mathrm{T}} 
\lesssim 0.3$) and using the relation from \citet{lfr+07}, we 
find $\log (M_{\mathrm{BH}} / M_{\odot}) \lesssim 7.5$ [$\log 
(M_{\mathrm{BH}} / M_{\odot}) \lesssim 7.0$, neglecting any 
host-galaxy extinction correction], suggesting a similar mass black hole to
that found in the centre of the Milky Way ($M_{\mathrm{BH}} \approx
4 \times 10^{6}$\,M$_{\odot}$; \citealt{gsw+08}).  

Finally, we can use the results of our template fitting to infer
additional global properties of the galaxy \hosts.  Following
\citet{khw+03}, we calculate the strength of the 4000\,\AA\ break
(D$_{n}$(4000) $= 1.2 \pm 0.1$) and the equivalent width (EW) of the
H$\delta$ absorption feature (EW(H$\delta$) $= 3.9 \pm 1.0$\,\AA).  
 The ratio of D$_{n}$(4000) to EW(H$\delta$) indicates that \hosts\ has 
been undergoing relatively continuous star formation over its history 
(see Figure~3 from \citealt{khw+03}).  However, the
relatively strong H$\delta$ feature suggests a significant
contribution from A-type stars, and we infer that the bulk of star
formation likely ended within the last 0.1--1\,Gyr.
The current star-formation rate, derived from the
extinction-corrected luminosity in H$\alpha$, is $1.5 \pm
0.5$\,M$_{\odot}$\,yr$^{-1}$ (using the calibration from
\citealt{k98}).

We can further utilise the derived 4000\,\AA\ break strength and
H$\delta$ EW to estimate the mass-to-light ratio in
\hosts.  Using the models from \citet{khw+03}, we find $-0.4 \lesssim
\log (M / L) \lesssim -0.1$.  We thus infer a total stellar mass of
$M \approx 10^{10}$\,M$_{\odot}$ from our extinction-corrected
spectrum.  

\section{Interpretation}
\label{sec:interp}

Having completed our analysis of the available data
(\S\ref{sec:analysis}), we now turn our attention to interpretation.
Ultimately, we wish to understand the physical processes responsible
for the observed transient emission from PTF10iya.  Given the
extreme luminosity and short-lived duration, here we consider four
potential origins for PTF10iya: the afterglow of a gamma-ray burst
(GRB; \S\ref{sec:grb}), a core-collapse supernova (\S\ref{sec:sn}), 
gas accretion onto a central SMBH (AGN or LINER;
\S\ref{sec:agn}), and the tidal disruption of a star by an (otherwise
quiescent) SMBH (\S\ref{sec:tdf}).

\subsection{Gamma-ray burst afterglow}
\label{sec:grb}
For their brief durations, GRBs are the most luminous
transients in the Universe, reaching peak optical absolute magnitudes as large
as $M_{V} \approx -37$\,mag (GRB\,080319B; \citealt{bpl+09,rks+08})
and X-ray luminosities well in excess of $10^{44}$\,erg\,s$^{-1}$
(e.g., \citealt{ebp+07}).  However, GRB afterglow emission, 
widely believed to result from synchrotron radiation from electrons 
accelerated by an outgoing relativistic blast wave (see, e.g., \citealt{p05}), 
is a distinctly {\it nonthermal} process.  The SED in the optical
regime is typically well-described by a power law ($f_{\nu} \propto
\nu^{-\beta}$) with $\beta \approx 0.5$ to $1.5$ (i.e., $f_{\nu}$
increases at longer wavelengths; \citealt{spn98}).  Extinction in the
host galaxy only serves to make the observed spectrum even redder.
This contrasts sharply with the observed UV/optical emission from PTF10iya.  
Even if the SED of PTF\,10iya does not match a blackbody, it is 
much too blue to be compatible with a GRB afterglow.

\subsection{Supernova}
\label{sec:sn}
Like GRB afterglows, core-collapse SNe are capable 
of generating bright, transient X-ray (e.g., $L_{\rm X} = 6 \times
10^{43}$\,erg\,s$^{-1}$; \citealt{sbp+08,mlb+09b}) and optical (e.g., $M =
-22.7$\,mag; \citealt{qaw+07}) emission.  Furthermore, the spectra of
young core-collapse SNe are typically dominated by a blue,
thermal continuum.

However, PTF10iya is unlikely to have resulted from any known class of 
SN.  To begin with, the derived blackbody radius ($\sim 300$\,AU) is
much larger than the radius of a typical stellar progenitor, and
the outgoing ejecta would not have time to reach such a large distance 
traveling at typical SN velocities.  Furthermore, all
known X-ray SNe are at least two orders of magnitude fainter
than PTF10iya.\footnote{The X-ray emission observed from SN\,2008D
  resulted from the breakout of the shock from the stellar envelope
  \citep{c74}.  As such, it was quite short-lived (hundreds of
  seconds), and disappeared entirely as late as 5\,d after the
  detected outburst.}  The X-ray emission from core-collapse
SNe typically results from shock heating of a moderately dense
circumstellar medium (e.g., \citealt{il03}).  It is therefore quite
long-lived, in some cases for decades. Moreover, the optical spectra
of such SNe generally exhibit narrow and intermediate-width
emission lines (e.g., \citealt{scs+10}).

Finally, we note that the decay timescale observed in the optical 
($\sim 0.3$\,mag\,d$^{-1}$) is shorter than that of any 
previously observed SNe thought to be powered
by radioactive decay.\footnote{The fast decay rate could result to
  some extent from spectral evolution (i.e., varying
  $T_{\mathrm{BB}}$), and therefore may not represent the true
  evolution of the bolometric luminosity.  For example, the light
  curve of PTF\,09uj, which is believed to be powered by shock
  breakout from a dense circumstellar wind, declined even
  more quickly in the $R$ band ($\sim 3$\,mag in 25 days;
  \citealt{orn+10}).  We have already, however, dismissed a 
  shock-breakout origin for PTF10iya.}  The current record holders, 
SN\,2010X \citep{kkg+10} and SN\,2002bj \citep{dcn+10},
both decayed after the peak at a rate of $\sim 0.2$\,mag\,d$^{-1}$.
It would require an extremely small ejecta mass
and/or an outflow entirely transparent to the $\gamma$-rays produced by
the radioactive decay to produce a SN that decayed at such a
rapid rate, both of which are inconsistent with the large observed
luminosity.   

\subsection{AGN/LINER}
\label{sec:agn}

The nuclear location of PTF10iya, together with the observed
spectrum (reminiscent of the ``big blue bump''; \citealt{spn+89,b90b}), 
naturally lead us to consider accretion onto a SMBH
as a possible origin \citep{s78,ms82}.  Here we determine if 
gas accretion (either in the form of a normal AGN or a LINER) 
is consistent with the observed properties of PTF10iya.

To begin with, we must confront the fact that the nebular emission
lines in the spectrum of \hosts\ do not appear to be caused by
AGN photoionisation, even with a low ionisation parameter as in
a LINER. In and of itself, however, this is not sufficient to
entirely discount continuous accretion activity.  Because of the
nonzero size of our slit, our spectrum of the ``nucleus'' of
\hosts\ could be contaminated by nearby H\,II regions.  For example, 
{\it Hubble Space Telescope (HST)} spectroscopy of NGC\,5905, initially
classified as a starburst galaxy in a similar analysis to that
conducted in \S\ref{sec:host}, revealed a faint Seyfert nucleus that
had previously gone undetected \citep{ghk+03}.

If we set aside for the moment the nebular emission-line
classification, we still must simultaneously account for the short
timescale and the order of magnitude increase in the UV/optical and
X-ray flux.  The most dramatically variable class of AGNs are the 
blazars, which are thought to arise when a relativistic jet fed by 
accretion is oriented directly along our line of sight (e.g.,
\citealt{br78,up95}).  For example, 3C\,279 has varied across the entire
electromagnetic spectrum by factors of order a few on timescales
shorter than days (e.g., \citealt{wpu+98}).

Blazars are typically hosted in massive 
($\gtrsim 10^{10}$\,M$_{\odot}$) elliptical galaxies, though
individual counterexamples are known (see, e.g., \citealt{gof+02}, for
a blazar initially thought to be an orphan GRB afterglow).  More
importantly, blazars are almost exclusively bright radio sources, even
in quiescence.  Using our post-outburst radio limits from the EVLA
(\S\ref{sec:obs:radio}), we calculate an upper limit on the
quiescent radio luminosity ($\nu L_{\nu}$) of $< 3 \times
10^{38}$\,erg\,s$^{-1}$ at $\nu = 1.3$\,GHz.  Even when comparing with
samples of blazars selected based on X-ray variability, the ratio of
quiescent optical to X-ray flux ($\alpha_{\rm OX} \equiv -\log (f_{\nu,{\rm O}}
/ f_{\nu,{\rm X}}) / \log ( \nu_{\rm O} / \nu_{\rm X} ) > 1.6$), like the ratio of 
quiescent radio to optical flux ($\alpha_{\rm RO} \equiv -\log (f_{\nu,{\rm R}}
/ f_{\nu,{\rm O}}) / \log ( \nu_{\rm R} / \nu_{\rm O} ) < 0.1$), appears 
inconsistent with the properties of the known blazar population 
($\alpha_{\rm OX} = 0.6$--1.4; $\alpha_{\rm RO} = 0.2$--0.5; 
\citealt{beb+03}).

Even during outburst, the characteristic double-peaked blazar SED
(e.g., \citealt{up95,fcg+97}) does not provide a good match to the
observed emission from PTF10iya.  Well below the first peak ($\nu
L_{\nu}$) in the broadband spectrum, blazar SEDs are dominated by
nonthermal synchrotron radiation, rising as a power law like
$\nu^{1}$.  The SED only becomes shallower near the peak frequency.
The observed spectrum of PTF10iya, on the other hand, is quite steep,
with $\nu L_{\nu} \propto \nu^{2.7}$.  Furthermore, the SED of lower
luminosity blazars typically peaks at higher frequencies than that of 
higher luminosity sources (e.g., \citealt{fmc+98}).  Objects with peak
frequencies as low as the UV typically have peak $\nu L_{\nu} \gtrsim
10^{45}$\,erg\,s$^{-1}$, several orders of magnitude brighter than
PTF\,10iya.  In contrast, the faintest blazars compiled in the sample
of \citet{fmc+98}, with peak $\nu L_{\nu} \approx 6 \times
10^{44}$\,erg\,s$^{-1}$ (a factor of several brighter than
PTF10iya), peak in the soft X-ray band. 

Turning now to Seyfert galaxies, we consider two separate classes of
known sources, and compare their variability properties to those of PTF10iya.
{\it ROSAT} detected bright ($L_{\rm X} \gtrsim 10^{42}$\,erg\,s$^{-1}$) X-ray
outbursts from a number of narrow-line Seyfert 1 (NLS1) galaxies as
part of the All-Sky Survey \citep{vab+99}.  For instance, the nearby
($z = 0.028$) galaxy WPVS\,007 was detected in outburst with $L_{\rm X}
\approx 10^{44}$\,erg\,s$^{-1}$ in November 1990; three years later,
the X-ray flux had declined by a factor of 400 \citep{gbm+95}.  While
WPVS\,007 is the most dramatic example, many other NLS1 galaxies were
observed by {\it ROSAT} to vary by more than an order of magnitude on 
timescales as short as days (e.g., \citealt{mpu+04}).  In some cases, the 
X-ray flare was followed by variability in the optical spectra, including
the appearance of high-ionisation iron emission lines (e.g., IC\,3599;
\citealt{gbm+95,bpf95,kb99}).

However, several properties of either PTF\,10iya (the outburst)
or \hosts\ (the quiescent host) are inconsistent with the NLS1
interpretation.  To begin with, the X-ray spectra of NLS1 galaxies are
typically significantly steeper (power-law index $\Gamma \approx
3$--5 compared to $\Gamma_{\mathrm{PTF10iya}} = 1.8$) than the 
observed X-ray outburst from PTF10iya \citep{gwl+04}.
The X-ray spectral slope is actually quite crucial to the
interpretation of the NLS1 phenomenon, where the observed variability
is thought to arise from changes in the covering fraction of a warm
absorbing cloud \citep{glk08}.  In addition, optical spectra
of NLS1 galaxies are characterised by (1) relatively narrow Balmer
emission lines (FWHM(H$\beta$) $< 2000$\,km\,s$^{-1}$); (2) weak
[O\,III] $\lambda$5007 emission ([O\,III]/H$\beta < 3$),
and (3) strong Fe\,II emission \citep{op85,g89}.  For \hosts, the
[O\,III] $\lambda$5007 emission is indeed quite weak.  But while
the H$\alpha$ and H$\beta$ emission lines are clearly narrow, there is
in fact no sign of a broad component whatsoever --- both H$\beta$ and
H$\alpha$ are unresolved in our July LRIS spectrum.  Together with the
lack of Fe\,II emission, PTF10iya appears to differ fundamentally
from the known properties of NLS1 galaxies, or of normal Seyfert 1
galaxies for that matter.

While a Seyfert 2 galaxy (those lacking any evidence for a broad-line
region in the optical) may be easier to hide in the spectrum of the
quiescent host galaxy, dramatic continuum variability is not typically
observed from these sources.  One particularly interesting exception 
merits mention here: NGC\,5905 \citep{bkd96,kb99}.  

Bright X-ray emission ($L_{\rm X} = 7 \times 10^{43}$\,erg\,s$^{-1}$) was
detected from NGC\,5905 in 1990 July by the {\it ROSAT} all-sky survey.  
Follow-up observations over the subsequent months and years failed to
redetect the source, implying that the quiescent flux is at least 
two orders of magnitude fainter than the outburst.  Like in the NLS1
galaxies, the X-ray spectrum during outburst was quite soft ($\Gamma
\approx 4$--5).  But follow-up optical spectroscopy, both from the ground and
with \textit{HST}, failed to reveal any broad component to the Balmer
emission lines \citep{kb99,ghk+03}.  

A variety of models have been proposed to explain the properties of
the NGC\,5905 outburst (e.g., \citealt{kb99,lnm02}).  In the AGN
context, \citet{kb99} explored the possibility of a variable absorbing
cloud along the line of sight.  In this warm absorber model, the
Seyfert nucleus becomes visible by ionizing the ambient medium, making
it transparent to soft X-rays (much like a NLS1).  
This naturally explains the relatively
steep X-ray spectrum observed in NGC\,5905 ($\Gamma \approx 5$), but
the presence of a significant dust column is required to hide the
broad-line region in the optical.  

PTF10iya is more difficult to accomodate in the warm absorber
framework, as the observed X-ray spectrum is not particularly steep
($\Gamma = 1.8$), and the observed thermal SED in outburst rules
out a large amount of dust (at least at that time).  We cannot,
however, entirely eliminate geometries whereby the observed outburst 
results from short-lived changes in obscuration along the line of
sight.  The lingering ambiguity over the origin of the NGC\,5905
outburst (see, e.g., \citealt{lnm02} for a tidal disruption flare
interpretation) reinforces just how difficult such a task can be.

Finally, we consider the known variability properties of 
low-luminosity AGNs (i.e., LINERs) to see if they could explain 
our observations of PTF10iya.  It was
originally thought LINERs exhibited little to no X-ray variability on
short timescales (e.g., \citealt{kbh99}), though more recent
studies of Type I LINERs
have demonstrated variations of $\sim 30$\% on timescales of less
than one day \citep{prm+10}.  Likewise, recent UV studies have shown
variability of order 50\% in the UV in both Type I and II LINERs, though
on longer (months) timescales \citep{mnf+05}.  In fact, at the time
of our X-ray detection of PTF10iya (2010 June 11), the observed
optical to X-ray spectral index, $\alpha_{\rm OX} \approx 1.1$, is broadly
consistent with the properties of known low-luminosity AGNs
\citep{m07}. 

However, if \hosts\ does indeed host a faint active nucleus (similar
to NGC\,5905; \citealt{ghk+03}), the emission-line flux from the
nucleus must be significantly less than the total integrated value
across the slit (in order for the line diagnostics to indicate a pure 
H\,II region).  If we adopt a limit of $< 10\%$ on the contribution to
the extinction-corrected [O\,III] line and apply the observed
correlation between [O\,III] flux and $L_{\rm X}$ for AGNs \citep{hph+05},
we infer a quiescent X-ray luminosity of $L_{\rm X} \lesssim
10^{42}$\,erg\,s$^{-1}$.  This value is two orders of magnitude less
than the observed flaring state; to the best of
our knowledge, such a dramatic outburst has never before been observed
in a low-luminosity LINER, though the nature of the variability in 
these sources is not yet well understood.

To summarise, even if we overlook the lack of evidence for AGN
activity in \hosts, the emission-line ratios, the historical light
curve, and the dramatic increase in the X-ray flux make PTF10iya unique
among known AGN outbursts.
Of course, we cannot entirely exclude some previously unobserved mode of
variability in the gas accretion process.  Nevertheless, given the
inadequacy of AGNs in explaining the observed properties 
of PTF10iya, we also seek an alternative possibility.

\subsection{Tidal disruption flare}
\label{sec:tdf}

Stars passing too close to a SMBH will be
torn apart when tidal forces are stronger than their self-gravity
\citep{fr76,lth82,r88}.  In the case of solar-type stars (mass $M_{*}$, radius
$R_{*}$), this disruption will occur outside the event horizon for
SMBHs with $M_{\mathrm{BH}} \lesssim 10^{8}$\,M$_{\odot}$
\citep{h75}.  Bound stellar debris is sent off on highly elliptical
orbits, and should return to pericentre ($R_{\mathrm{p}}$) at a rate
$\dot{M} \propto t^{-5/3}$ \citep{p89,ek89,alp00,lkp09,rr09}.  
The resulting accretion onto the SMBH powers an outburst known 
as a tidal disruption flare (TDF).

In the simplest models, the returning debris
shocks and circularises, forming a torus around the SMBH (e.g.,
\citealt{u99,lnm02}).  Particularly at late times, the disc should be
optically thick but geometrically thin, and the emergent spectrum
should be approximately thermal, with
\begin{eqnarray}
T \approx \left ( \frac{L_{\mathrm{Edd}}} {4 \pi R_{\mathrm{T}}^{2}
  \sigma} \right )^{1/4} = \nonumber \\
  2 \times 10^{5} \left (
  \frac{M_{\mathrm{BH}}}{10^{6}\, {\rm M}_{\odot}} \right )^{1/12} \left ( 
  \frac{R_{*}}{{\rm R}_{\odot}} \right )^{-1/2} \left ( 
  \frac{M_{*}}{{\rm M}_{\odot}} \right )^{-1/6} \mathrm{K},
\label{eqn:Tdisk}
\end{eqnarray} 
where $R_{\mathrm{T}} \approx R_{*} (M_{\mathrm{BH}} / M_{*})^{1/3}$
is the tidal radius (where the tidal force due to the SMBH exceeds the
self-gravity of the star).  As is evident from Equation~\ref{eqn:Tdisk}, the
spectrum peaks in the far UV, with a large bolometric correction from
the rest-frame optical bandpass.  For emission from the accretion
disc, the bolometric luminosity evolution should approximately follow
the mass accretion rate, $L \propto t^{-5/3}$, although the observed
flux in a given bandpass could deviate from this behaviour.

It is not surprising, then, that the majority of TDF candidates
have been identified through wide-field X-ray
\citep{bkd96,kg99,gsz+00,dbe+02,esf+07,esk+08,mue10} and UV
\citep{rgd+95,gmm+06,gbm+08,ghc+09} surveys.  PTF10iya differs in
important ways from most of these TDF candidates: the derived
blackbody temperature of PTF10iya is approximately an order of
magnitude smaller than in these sources (which exhibit 
typical temperatures of a few times
$10^{5}$\,K), the decay timescale is significantly shorter (typical
durations of months to years), and, when detected, the X-ray
emission is significantly harder (typical power-law indices of $\Gamma
\approx 4$--5).  Many of these differences could be ascribed to
bandpass selection effects, so a more apt comparison would be to the
{\it optically} discovered candidates from \citet{vfg+10}.  We
shall return to this issue shortly.

The blackbody prediction in Equation~\ref{eqn:Tdisk} is well motivated
at late times, when the fallback rate is sub-Eddington: 
$\dot{M} \lesssim \dot{M}_{\mathrm{Edd}}
\equiv L_{\mathrm{Edd}} / \epsilon c^{2}$ ($\epsilon$ is the radiative
efficiency of the accretion disc, assumed to be $\sim 0.1$).
However, for black holes with $M_{\mathrm{BH}} \lesssim 3 \times
10^{7}$\,M$_{\odot}$, the mass accretion rate will {\it exceed} the
Eddington rate for some period of time following disruption.  In this
case, the   
accretion disc is likely to be highly advective, and some fraction of
the bound material may be blown out from the system as a powerful wind
(e.g., \citealt{kp03,omn+05}).  \citet{sq09} have calculated the expected emission
from this super-Eddington outflow, and predict a short-lived ($\Delta
t \approx 30$\,d), luminous ($L_{\mathrm{peak}} \approx
10^{44}$\,erg\,s$^{-1}$) flare with a lower temperature than
that seen during the thin-disc phase.

The observed properties of PTF10iya are broadly consistent with the
predictions for this super-Eddington phase.  The
relatively short timescale, both in terms of the rise and decline,
suggests that the disruption occurred close to the SMBH:
$R_{\mathrm{p}} \lesssim 3 R_{\mathrm{S}}$, where $R_{\mathrm{S}}$ is
the Schwarzschild radius.  Likewise, the large bolometric luminosity
is consistent with a relatively massive SMBH: $M_{\mathrm{BH}}
\gtrsim 10^{7}$\,M$_{\odot}$.  For a solar-type star with
pericentre distance $R_{\mathrm{p}} \approx 2 R_{\mathrm{s}}$
disrupted by a $10^{7}$\,M$_{\odot}$ black hole, we find
$R_{\mathrm{BB}} \approx 100$\,AU, $T_{\mathrm{BB}} \approx
10^{4}$\,K, and $L_{\mathrm{BB}} \approx 10^{45}$\,erg\,s$^{-1}$ a few
days after the disruption occurs [taking the fiducial values for the
fraction of incoming material expelled in the super-Eddington wind
($f_{\mathrm{out}}$) and the terminal wind velocity
($f_{v}$) from \citealt{sq09}].  If we instead allow
$f_{\mathrm{out}}$ and $f_{v}$ to vary, we find a range of possible
models ($M_{\mathrm{BH}} \approx 10^{6}$--$10^{7}$\,M$_{\odot}$;
$R_{\mathrm{p}} \approx 2 R_{\mathrm{S}}$--$R_{\mathrm{T}}$) can
reproduce the observed optical luminosity and colours.
With only two epochs of data,  however, a detailed test of these 
models, in particular the temporal evolution, is beyond the scope 
of this work.  

After the accretion rate drops below Eddington, in the tidal
disruption model the emission should be dominated by shock-heated
material in the accretion disc.  Given the predicted temperature for
the disc in this phase (Eq.~\ref{eqn:Tdisk}), it is important to
check if we would be sensitive to any disc emission in our X-ray and
UV observations from 2010 August.  Using models for the sub-Eddington
disc emission from \citet{sq09} with the system parameters determined
above, we estimate $\nu L_{\nu} ( \lambda_{\mathrm{rest}} =
2000$\,\AA$) \lesssim 2 \times 10^{42}$\,erg\,s$^{-1}$.  This
corresponds to a UVOT $UVW1$ magnitude of $\gtrsim 22.5$.  Even
neglecting the derived host-galaxy extinction, this value is several
magnitudes fainter than our limits at this time.  Likewise, the
derived X-ray emission from this accretion disc component is well
below the upper limits from the XRT at this time.

On the other hand, comparing our late-time ($\Delta t \approx
  2$\,months) limits with the {\it observed} UV properties of
  candidate TDFs from GALEX presents a slightly more complex picture.
  With typical near-UV absolute magnitudes (AB) of $\approx -18$ to
  $-19$\,mag at this time \citep{ghc+09}, a comparable source would 
  appear with $UVW1 \approx 19.5$--$20.5$\,mag at the distance 
  of PTF10iya.  One possibility to explain this apparent discrepancy
  with our derived limits ($UVW1 > 20.5$\,mag) is extinction: our 
  derived value of $E(B-V)_{\mathrm{gas}} = 0.40$\,mag corresponds to 
  $A_{UVW1} \approx 1.5$\,mag.  With this correction, our 2010 August 
  UVOT limits are not sufficiently deep to rule out UV flares similar
  to those reported by \citet{ghc+09}.  However, our derived
  extinction correction is quite uncertain.  We are therefore left to 
  conclude that our 2010 August \Swift\ observations are not 
  sufficiently constraining to either detect or rule out the expected 
  disc signature during the sub-Eddington phase.

A $10^{7}$\,M$_{\odot}$ SMBH is consistent with the limits for
\hosts\ derived from the black hole vs. bulge luminosity relation
(\S\ref{sec:host}).  Furthermore, \citet{sq10} predict that any absorption
features in the optical bandpass due to the photoionisation of
the unbound material should be weaker for more massive black holes,
and indeed we do not detect such features in the early-time 
spectra of PTF10iya.  We reiterate, however, that
the estimated black hole mass is still sufficiently small to ensure
that the disruption would occur outside the event horizon.

In addition to the outburst properties, the host galaxy of PTF10iya is
consistent with a TDF interpretation.  No variability is seen in
the optical light curve in our PTF and DeepSky imaging of the field in
the last few years.  The observed emission-line ratios in quiescence 
indicate a moderately star-forming galaxy and are incompatible with
photoionisation by the hard X-ray continuum common in AGNs.  In fact, 
the host galaxy of PTF10iya, \hosts, bears a strong
resemblance to the host galaxy of D23H-1, a candidate TDF
discovered by \textit{GALEX} \citep{ghc+09}.  The diagnostic
emission-line ratios are all consistent within 2$\sigma$
uncertainties, as is the derived line-of-sight extinction.  

Several important distinctions between PTF10iya and previous TDF
candidates remain, however.  First, PTF10iya appears to decay much
more rapidly than any previously claimed TDF candidate, even the
handful discovered in the optical.  Given that
the constraints on the outburst dates are much weaker for nearly all
previous candidates\footnote{Two exceptions include D1-9, where the
  a rising light curve was observed in the optical \citep{gbm+08}, and
  NGC\,5905, where a rising X-ray light curve was observed
  \citep{lnm02}.}, this may be simply because PTF10iya was caught
shortly after the disruption; the higher cadence of PTF enables
detection of short-lived outbursts that might have been missed by
other surveys.  Nevertheless, the contrast between the
two TDF candidates from \citet{vfg+10}, which decayed by $\sim
0.5$--1.0\,mag in the 50\,d since discovery, and PTF10iya, which
decayed by $\sim 1.5$\,mag in only 6\,d, is quite striking.

Most importantly, the X-ray emission from PTF10iya, which accounts for
a significant fraction of the bolometric luminosity, remains a
significant challenge for TDF models.  Unlike previous TDF candidates
discovered in the X-rays, the shallow, nonthermal X-ray spectrum is
inconsistent with a simple extrapolation of a $10^{5}$\,K blackbody to
high energies.  

Recently, two unusual high-energy transients discovered by the
  \Swift\ satellite (\textit{Sw}\,J1644+57 and \textit{Sw}\,J2058+05)
  were suggested to result from the tidal disruption process
  \citep{ltc+11,bgm+11,bkg+11,zbs+11,ckh+11b}.  While both sources
  display extremely luminous ($L_{\rm X} > 10^{47}$\,erg\,s$^{-1}$) X-ray
  emission, this was also accompanied by a luminous radio source, with
  compelling evidence that the outburst marked the birth of a
  collimated, relativistic outflow (i.e., a jet).  One possibility is
  that the observed emission from PTF10iya is generated by the same
  process (i.e., tidal disruption), but we are viewing the system from
  outside the narrow angle of the jet (an ``off-axis'' TDF).
  
\citet{gm11} considered off-axis emission from the interaction of a
  mildly relativistic jet with surrounding material, and predict
  bright ($L_{\nu} \gtrsim 10^{30}$\,erg\,s$^{-1}$\,Hz$^{-1}$ at 10
  GHz), self-absorbed radio emission peaking around a year after
  disruption\footnote{\citet{vkf+11} also consider radio jets
    from TDFs, but powered instead internally by the accretion
    process.  The predicted radio luminosity and spectra are mostly
    similar to those of \citet{gm11}.}.  Our radio nondetections
  (\S\ref{sec:obs:radio}) are an order of magnitude below the
  predictions of these models, suggesting in this case either 
  that no relativistic jet was generated during the disruption 
  process or that the surrounding environment differed significantly from
  the nominal parameters assumed by \citet{gm11}.  We note that these
  limits are significantly more constraining for these radio jet
  models than previous observations, which were either obtained
  shortly after disruption \citep{vfg+10} or solely at low frequency,
  where self-absorption comes into play \citep{b11}.

If we do not consider relativistic jetted models, we may still be able
to explain the observed X-ray flare by analogy with AGNs.  
Theoretically, for $M_{\mathrm{BH}} > 10^{7}$\,M$_{\odot}$, the
density at pericentre is never quite high
enough for the gas and radiation to reach thermal equilibrium, and so
instead, photons likely Compton upscatter off of fast-moving electrons
to produce hard X-rays \citep{sq10}.  While such a phenomenon, similar
to what is observed in the hard X-ray bandpass in AGNs, may account for
the X-ray emission from PTF10iya, the process is relatively poorly
understood and beyond the scope of this work.

\section{Conclusions}
\label{sec:conclusions}
To summarise, we reiterate the primary observed properties of
PTF10iya.
\begin{enumerate}
\item It was a short-lived ($\tau \approx 10$\,d), luminous ($M_{g^{\prime}}
  \approx -21$\,mag) UV/optical flare.
\item After subtracting the host-galaxy contribution, the transient emission
  is relatively blue, and reasonably well-fit by a blackbody with 
  $T \approx$ (1--2) $\times 10^{4}$\,K, $R \approx 200$\,AU, and $L
  \approx 10^{44}$--$10^{45}$\,erg\,s$^{-1}$.
\item A simultaneous X-ray flare of comparable energy output ($L_{\rm X}
  \approx 10^{44}$\,erg\,s$^{-1}$) at $\Delta t \approx 5$\,d after the
  outburst was detected.
\item It was astrometrically consistent with the nucleus of a $z = 0.22405$
  star-forming galaxy with a predominantly disc-like morphology (\hosts).
\item There is no evidence, either from the historical light curve or 
  host spectrum, for previous episodes of accretion onto the central 
  SMBH; that is, \hosts\ does not appear to be an AGN.
\end{enumerate}

After ruling out other known classes of extragalactic variable sources
(GRBs and SNe),
we have demonstrated that the basic properties of the UV/optical
outburst (luminosity, temperature, duration) are broadly consistent
with the recent predictions of \citet{sq09} for the early stages
following the tidal disruption of a solar-type star by a $\sim
10^{7}$\,M${_{\odot}}$ SMBH.  This picture is further supported by
the lack of variability in the historical optical light curve, and the
emission-line diagnostic diagrams which suggest an ordinary
star-forming galaxy (i.e., not a Seyfert or LINER).  

At the same time, PTF10iya differs in important ways from previously
identified TDF candidates.  While some of this can be understood
largely as bandpass biases (e.g., lower blackbody temperature than for
UV and X-ray selected candidates), the short duration and associated
bright X-ray flare suggest a significant diversity if indeed all of 
these sources do result from the tidal disruption of a star by a SMBH.
Without an unambiguous signature, such as the predicted broad
($0.01c$--$0.1c$), blueshifted absorption features in the rest-frame UV
bandpass \citep{sq10}, we cannot definitely rule out an unusual AGN
outburst scenario.

While much of the interpretation herein has been centred
on the TDF framework,
it is nonetheless interesting to speculate on what might cause such a
short timescale outburst in an otherwise quiescent AGN.  The
black-hole binary systems known as soft X-ray transients (SXTs)
at first glance appear to be potential stellar-mass analogs.  These
systems, such as A0620--00 \citep{epp+75,gho01}, spend decades in a
quiescent state accreting from their companions at extremely low rates
($\dot{M}_{\mathrm{Edd}} < 0.01$).  After intervals of several decades,
however, a thermal viscous instability in the accretion disc triggers
a dramatic outburst typically lasting several weeks (e.g.,
\citealt{c93}).  Since the recurrence time should scale proportionally
to black hole mass, a quiescent $10^{7}$\,M$_{\odot}$ black hole would
be expected to undergo such an outburst every $\sim 10^{6}$\,yr, much
too long to detect on human timescales.  Interpreting AGN outbursts
in this manner yields a self-consistent quasar luminosity function
\citep{se97}, and can account for the resemblance between the power
spectrum of AGNs and stellar-mass black-hole binaries in the soft-high
state \citep{mpu+04,mkk+06}.

There are significant problems with this picture, however.  First, 
the thermal viscous instability may not be capable of producing
large-amplitude outbursts in SMBH systems, as the gas in these 
systems is well coupled to the magnetic field even in quiescence due to
the larger size of the accretion disc \citep{mq01}.  Even more
troubling, the outburst duration should also scale as the black hole
mass, so SXT-like outbursts in SMBH systems would have durations of
$\sim 10^{4}$--10$^{5}$\,yr.  Conversely, a 10\,d outburst in a
$10^{7}$\,M$_{\odot}$ system corresponds to a $\sim 1$\,s outburst in
a 10\,M$_{\odot}$ black-hole binary.  While some stellar-mass Galactic
systems do exhibit dramatic X-ray flares on such short timescales (e.g.,
GRS\,1915+105; \citealt{gmr96,mrg97,mmr99}), these systems do not
return to quiescence immediately (if at all).  To the best of our
knowledge, there are no known stellar-mass black-hole binaries that
transition from quiescence to Eddington luminosities on such a short
timescale.  If PTF10iya is truly an AGN outburst, it would suggest
some new accretion physics that does not appear to manifest itself 
in stellar-mass systems.

Regardless of its ultimate origin, it is clear that events like 
PTF10iya have not been previously reported.
But because of its short timescale, this does not
necessarily mean such outbursts are rare.  Given the large luminosity
(comparable to the brightest SNe), current and future
optical surveys should be capable of detecting additional similar
events out to relatively large distances, provided they observe with
sufficiently high cadence.  Rapid broadband follow-up observations
will be key to uncovering more examples of this class (see, e.g.,
\citealt{gka+11}, for a detailed discussion of the PTF follow-up
time line).  

\section*{Acknowledgments}

We wish to thank S.~A.~Wright, K.~Bundy, and the anonymous referee
for valuable comments and discussions regarding this manuscript.  
Follow-up data were obtained by the Palomar Transient Factory TDF 
Key Project. 

S.B.C. and A.V.F. wish to acknowledge generous support from Gary and
Cynthia Bengier, the Richard and Rhoda Goldman Fund, National Aeronautics and
Space Administration (NASA)/\Swift\ grant NNX10AI21G, NASA/{\it Fermi} grant
NNX1OA057G, and National Science Foundation (NSF) grant AST--0908886.  
N.R.B. is supported through the
Einstein Fellowship Program (NASA Cooperative Agreement NNG06DO90A).
J.S.B. and his group were partially supported by NASA/\Swift\ Guest
Investigator grants NNX09AQ66G and NNX10AF93G, and a SciDAC grant 
from the US Department of Energy.
The Weizmann Institute PTF partnership is supported in part
by grants from the Israeli Science Foundation (ISF) to A.G.
Joint work by the Weizmann and Caltech groups is supported
by a grant from the Binational Science Foundation (BSF) to
A.G. and S.R.K., and A.G. acknowledges further support from an
EU/FP7 Marie Curie IRG fellowship.  L.B. is supported by the NSF 
under grants PHY--0551164 and AST--0707633.
P60 operations are funded in part by NASA through the
{\it Swift} Guest Investigator Program (grant NNG06GH61G).

We acknowledge the use of public data from the \Swift\ data archive. 
Some of the data presented herein were obtained at the W.M. Keck 
Observatory, which is operated as a scientific partnership among 
the California Institute of Technology, the University of California,
and NASA. The
Observatory was made possible by the generous financial 
support of the W.M. Keck Foundation.  The authors wish to 
recognize and acknowledge the very significant cultural role 
and reverence that the summit of Mauna Kea has always had 
within the indigenous Hawaiian community; we are most 
fortunate to have the opportunity to conduct observations 
from this mountain. 



\bibliographystyle{mn2e}





\input{tab4.tex}

\label{lastpage}

\end{document}

%% file: tab1.tex
\ProvidesFile{deepsky.tab.tex}

\begin{table}
  \centering
  \begin{minipage}{80mm}
  \caption{Historical DeepSky observations of \hosts}
  \begin{tabular}{lcc}
Date\footnote{UT at midpoint of exposure.} & Exposure Time 
& Magnitude\footnote{Instrumental magnitudes were determined 
using a 2\farcs5 (radius) aperture.  Photometric calibration was 
performed relative to the SDSS $i^{\prime}$ filter and are on the AB 
magnitude system \citep{og83}.  Magnitudes here have not been 
corrected for Galactic extinction ($E(B-V) = 0.010$\,mag; 
\citealt{sfd98}), nor for extinction in the host galaxy.} \\
(UT) & (s) & \\
  \hline
2007 May 6.486 & 60.4 & $19.84 \pm 0.31$ \\ 
2007 May 18.462 & 60.2 & $19.63 \pm 0.30$ \\
2007 May 18.464 & 60.4 & $19.72 \pm 0.31$ \\
2007 May 18.476 & 60.5 & $19.82 \pm 0.33$ \\
2007 May 18.477 & 60.4 & $19.73 \pm 0.31$ \\
2008 May 8.476 & 60.2 & $19.60 \pm 0.24$ \\
2008 May 8.477 & 60.4 & $19.41 \pm 0.27$ \\
2008 May 8.485 & 60.3 & $19.66 \pm 0.27$ \\
2008 May 8.486 & 60.3 & $19.30 \pm 0.40$ \\
2008 Jun 10.326 & 60.3 & $19.34 \pm 0.26$ \\
2008 Jun 10.368 & 60.3 & $19.61 \pm 0.32$ \\
2008 Jun 13.306 & 80.5 & $19.61 \pm 0.30$ \\
2008 Jun 13.351 & 80.3 & $19.42 \pm 0.29$ \\
2008 Jun 18.239 & 100.3 & $19.62 \pm 0.34$ \\
2008 Jul 2.286 & 100.3 & $19.41 \pm 0.29$ \\
2008 Jul 2.317 & 100.4 & $19.55 \pm 0.29$ \\
2008 Jul 30.194 & 100.3 & $19.29 \pm 0.23$ \\
2008 Jul 30.238 & 100.3 & $19.32 \pm 0.26$ \\
  \hline
  \end{tabular}
  \label{tab:deepsky}
  \end{minipage}
\end{table}

%% file: tab2.tex
\ProvidesFile{spec.tab.tex}

\begin{table*}
  \begin{minipage}{15cm}
  \centering
  \caption{Optical spectra of PTF10iya}
  \begin{tabular}{lcccc}
Date\footnote{UT at beginning of exposure.} & 
Telescope/Instrument & Wavelength Coverage & 
Exposure Time & Spectral Resolution \\
(UT) & & (\AA) & (s) & (\AA) \\
  \hline
Jun 8.43 & Keck I/LRIS (blue) & 3500--5600 & 450.0 & 7.0 \\
Jun 8.43 & Keck I/LRIS (red) & 5600--10100 & 370.0 & 6.5 \\
Jun 9.10 & WHT/ISIS (blue) & 3100--5300 & 600.0 & 3.4 \\
Jun 9.10 & WHT/ISIS (red) & 5300--9500 & 600.0 & 7.3 \\
Jun 12.49 & Keck I/LRIS (blue) & 3500--5600 & 600.0 & 4.0 \\
Jun 12.49 & Keck I/LRIS (red) & 5600--10100 & 600.0 & 6.5 \\
Jul 8.29 & Keck I/LRIS (blue) & 3500--5600 & 1200.0 & 4.0 \\
Jul 8.29 & Keck I/LRIS (red) & 5600--10100 & 1200.0 & 6.5 \\
  \hline
  \end{tabular}
  \end{minipage}
\label{tab:spec}
\end{table*}

%% file: tab3.tex
\ProvidesFile{bbfits.tab.tex}

\begin{table}
  \centering
  \begin{minipage}{9cm}
  \caption{Optical/UV blackbody SED fit parameters}
  \begin{tabular}{lcccc}
Dates & $E(B-V)_{\mathrm{gas}}$ & $T$ & $L_{\mathrm{BB}}$ & $R_{\mathrm{BB}}$ \\
& (mag) & ($10^{4}$\,K) & ($10^{44}$\,erg\,s$^{-1}$) & (AU) \\
  \hline
Jun 6--9 & 0.40 & $2.1 \pm 0.3$ & $4.8 \pm 1.4$ & $130 \pm 20$ \\
Jun 6--9 & 0 & $1.3 \pm 0.2$ & $0.8 \pm 0.2$ & $140 \pm 30$ \\
Jun 11--13 & 0.40 & $2.2 \pm 0.3$ & $1.4 \pm 0.3$ & $63 \pm 13$ \\
Jun 11--13 & 0 & $1.5 \pm 0.2$ & $0.27 \pm 0.09$ & $57 \pm 12$ \\
  \hline
  \end{tabular}
  \end{minipage}
\label{tab:bbfits}
\end{table}

%% file: tab4.tex
\ProvidesFile{optical.tab.tex}

\begin{table*}
  \centering
  \begin{minipage}{12cm}
  \caption{UV/optical observations of PTF10iya}
  \begin{tabular}{lcccc}
  \hline
Date\footnote{UT at midpoint of exposure/coadd.} &
Telescope/Instrument & Filter & Exposure Time & 
Magnitude\footnote{All reported magnitudes have had host-galaxy flux
  removed and represent light only from the transient PTF\,10iya.
  Magnitudes here have not been corrected for Galactic extinction
  ($E(B-V) = 0.010$\,mag; \citealt{sfd98}), nor for extinction in the
  host galaxy of the transient.  Observations in the
  $g^{\prime}$-, $r^{\prime}$-, and $i^{\prime}$-bands are reported on 
  the AB magnitude system \citep{og83}.  $U$- and $R$-band
  observations are referenced to Vega, while $UVW1$, $UVM2$, and
  $UVW2$ are on the UVOT photometric system \citep{pbp+08}.} \\
(UT) & & & (s) & \\
  \hline
2010 Jun 1.295 & P48 & $R$ & 60.0 & $> 20.62$ \\
2010 Jun 1.344 & P48 & $R$ & 60.0 & $> 20.53$ \\
2010 Jun 6.302 & P48 & $R$ & 60.0 & $19.77 \pm 0.10$ \\
2010 Jun 6.356 & P48 & $R$ & 60.0 & $19.91 \pm 0.11$ \\
2010 Jun 7.170 & P60 & $i^{\prime}$ & 120.0 & $20.24 \pm 0.20$ \\
2010 Jun 7.172 & P60 & $r^{\prime}$ &120.0 & $19.92 \pm 0.10$ \\
2010 Jun 7.174 & P60 & $g^{\prime}$ & 120.0 & $19.79 \pm 0.08$ \\
2010 Jun 11.335 & P48 & $R$ & 60.0 & $> 19.39$ \\
2010 Jun 11.779 & \Swift/UVOT & $UVW1$ & 314.5 & $19.73 \pm 0.27$ \\
2010 Jun 11.782 & \Swift/UVOT & $U$ & 157.0 & $19.31 \pm 0.18$ \\
2010 Jun 11.784 & \Swift/UVOT & $UVW2$ & 629.4 & $19.73 \pm 0.17$ \\
2010 Jun 11.788 & \Swift/UVOT & $UVM2$ & 1478.2 & $19.77 \pm 0.15$\\
2010 Jun 13.311 & P60 & $i^{\prime}$ & 180.0 & $22.14 \pm 0.35$ \\
2010 Jun 13.313 & P60 & $r^{\prime}$ & 120.0 & $21.62 \pm 0.28$ \\
2010 Jun 13.318 & P60 & $g^{\prime}$ & 120.0 & $21.46 \pm 0.28$ \\ 
2010 Jun 16.334 & P48 & $R$ & 60.0 & $> 20.75$ \\
2010 Jun 16.378 & P48 & $R$ & 60.0 & $> 20.60$ \\
2010 Jun 18.687 & P60 & $i^{\prime}$ & 1080.0 & $> 22.24$ \\
2010 Jun 18.689 & P60 & $r^{\prime}$ & 720.0 & $> 22.69$ \\
2010 Jun 18.689 & P60 & $g^{\prime}$ & 360.0 & $> 21.61$ \\
\textbf{2010 Aug 10.498} & \Swift/UVOT & $UVM2$ & 1771.4 & $> 21.01$ \\
\textbf{2010 Aug 10.499} & \Swift/UVOT & $U$ & 591.4 & $> 20.67$ \\
\textbf{2010 Aug 10.499} & \Swift/UVOT & $UVW1$ & 1179.0 & $> 20.47$ \\
\textbf{2010 Aug 10.499} & \Swift/UVOT & $UVW2$ & 1998.5 & $> 20.69$ \\
  \hline
  \end{tabular}
 \label{tab:optuv}
  \end{minipage}
\end{table*}